\documentclass[aps,prd
,preprint
,tightenlines
,showkeys
,nofootinbib
,onecolumn
,amsfonts,amsmath,amssymb
]{revtex4}
\usepackage[pdftex]{graphicx}

\usepackage{bm,framed}
\usepackage{float,color}
\usepackage{physics} 
\usepackage{rsfso}
\newcommand{\pa}{\partial}
\newcommand{\del}{\delta}
\newcommand{\bs}[1]{\boldsymbol{#1}}

\newcommand{\annor}[1]{\textcolor{black}{#1}}
\newcommand{\anno}[1]{#1}

\begin{document}

\title{Interferometry of black holes with Hawking radiation}
\author{Yasusada Nambu}
\email{nambu@gravity.phys.nagoya-u.ac.jp}
\affiliation{Department of Physics, Graduate School of Science, Nagoya
University, Chikusa, Nagoya 464-8602, Japan}

\author{Sousuke Noda}
\email{snoda@cc.miyakonojo-nct.ac.jp}
\affiliation{National Institute of Technology, Miyakonojo College,
  Miyakonojo 885-8567, Japan} 
 \affiliation{
   Center for Gravitation and Cosmology, College of Physical Science
   and Technology,  Yangzhou University, Yangzhou 225009, China}

% 
%
%\affiliation{Department of Physics, Graduate School of Science, Nagoya
%University, Chikusa-ku, Nagoya 464-8602, Japan}
 
%
%\date{August 15, 2021 ver 0.3}
%\date{August 16, 2021 ver 0.4} 
%\date{August 23, 2021 ver 0.5}
%\date{August 28, 2021 ver 0.6}
% \date{August 31, 2021 ver 0.7}
% \date{September 6, 2021 ver 0.8}
%\date{September 8, 2021 ver 0.9}
%\date{September 14, 2021 ver 1.0} % submit version
% \date{December 13, 2021 ver 1.5} % resubmit version
% \date{December 18, 2021 ver 1.8} % after proofreading. resubmit version
%\date{January 30, 2022 ver 2.0} % after proofreading. resubmit
%version
\date{Februay 3, 2022 ver 2.0} % accepted version in PRD

\begin{abstract}
  We investigate the wave optical imaging of black holes with Hawking
  radiation. The spatial correlation function of Hawking radiation is
  expressed in terms of transmission and reflection coefficients for
  scalar wave modes and evaluated by numerically summing over angular
  quantum numbers for the Unruh-Hawking state of the
  Kerr-de Sitter black hole.  Then, wave optical images of an evaporating
  black hole are obtained by the Fourier transformation of the spatial
  correlation function.  For short wavelength, the image of the
  black hole with the outgoing mode of the Unruh-Hawking state has the
  appearance of  a star with its surface given by the photon sphere. It is
  found that interference between incoming modes from the cosmological
  horizon and reflected modes due to the scattering of the black hole
  can enhance brightness of images in the vicinity of the photon
  sphere. For a long wavelength, the entire field of view is bright, and
  the emission region of Hawking radiation cannot be identified.
\end{abstract} 

%\pacs{04.62.+v }

\maketitle

\tableofcontents

%%%%%%%%%%%%%%%%%%%%%%%%%%%%%%%%%%%%%%%%%%%%%%%%%%%
\section{Introduction}

In general relativity, a black hole is defined as a spacetime region
surrounded by the event horizon, from which no light signals can
escape to the null infinity.  There are several possibilities for the
formation of black holes in our universe: the gravitational collapse
of stars, coalescence of compact stars, phase transition in the early
universe, and so on. An important concept characterizing black holes
is the photon sphere, which is defined as a surface composed of
bounded null geodesics~\cite{Fro, Fro2, Teo}. \anno{For a rotating
  black hole, there are two circular photon orbits in the equatorial
  plane, and their radii differ depending on the signs of the photon
  angular momentum. There are also other bounded photon orbits, which
  depart from the equatorial plane\footnote{\anno{The radius of a
      bounded photon orbit is determined by combination of parameters
      $(L_z/E, C/E^2)$, where $C$ is the Carter constant, $L_z$ is the
      $z$ component of the angular momentum and $E$ is the energy
      of photon~\cite{Fro2}. The orbit forms a shell-like structure
      (photon shell)~\cite{Teo} .}}. We call a set of bounded photon
  orbits as a ``photon sphere'' in this paper.  When considering the
  propagation of ingoing null rays towards a black hole, null rays
  that cross the photon sphere cannot escape the photon sphere. From a
  distant observer, a set of projected bounded photon orbits on a far
  observer's view plane appears as a distorted disk that corresponds
  to the black hole shadow for a rotating black hole~\cite{Fro2}.}

Astrophysical black holes are associated with the 
surrounding gases showing light emission. Thus, the photon sphere of a
black hole can be visible as the rim of a dark shadow region in  bright
background emission. Indeed, recent observation of the central region
of M87 with very large baseline interferometry (VLBI) reported an
image of the photon sphere associated with the central supermassive
black
hole~\cite{Akiyama2019f,Akiyama2019h,Akiyama2019g,Akiyama2019k,Akiyama2019j,Akiyama2019i}.
As the apparent angular sizes of black hole candidates are very small from
the Earth, the key technology to resolve black hole shadows by
observation is aperture synthesis; by combining several independent
telescopes on the Earth, the effective size of the aperture can be
increased, making it possible to resolve black holes with very small apparent
sizes. The image reconstruction of black holes is performed 
based on a property of the wave optics known as the van
Cittert-Zernike theorem~\cite{Born1999,Wolf2007,Sharma2006}, which
states that the Fourier transformation of the first-order degrees of
coherence (interferometic fringe pattern) in an observer's screen
provides an intensity distribution of a source object if the spatial
incoherence of the source field is assumed.

In this paper, we aim to obtain wave optical images of evaporating
black holes. Owing to the quantum effect, black holes can emit thermal
radiation known as Hawking radiation~\cite{Hawking1974,Hawking1975a},
the temperature of which is proportional to the surface gravity of
black hole horizons. Hence, if we detect Hawking radiation of a black
hole from a spatially distant region, it is possible to reconstruct
wave optical images of the evaporating black hole by applying the van
Cittert-Zernike theorem. Of course, this investigation is only a
theoretical thought experiment because the Hawking temperatures of
astrophysical black holes are too low to detect directly. However, we
expect that our analysis will provide a deeper understanding of
Hawking radiation and black hole spacetimes from the viewpoint of wave
optics. In particular, it may be possible to acquire information on
the emission region of Hawking radiation by performing the wave
optical imaging of a black hole, and this direction of investigation
is related to the question ``where does Hawking radiation
originate?''~\cite{Giddings2016,Dey2017a}.  In our previous
studies~\cite{Kanai2013,Nambu2016}, we discussed the wave optical
imaging of black holes with a coherent point wave source.
Interference fringes due to wave scattering by a black hole appear on
the observer's screen. By the Fourier transformation of the
interference fringe, images of the Einstein ring and photon sphere are
obtained. For the case of an evaporating black hole, the wave source
is the black hole itself, and all the information necessary for
imaging is contained in the correlation function of Hawking radiation.
Concerning the quantum state of black holes, we assume the
Unruh-Hawking vacuum state, which is realized as black hole formation
via gravitational collapse
~\cite{Hawking1974,Hawking1975a,Unruh1976,Sataloff,Ottewill2000,Gregory2021,Fro,Birrell1984}.

\anno{In our analysis, instead of treating the asymptotically flat
  Kerr spacetime, we consider the Kerr-de Sitter (KdS) spacetime
  because it allows the evaluation of the reflection and transmission
  coefficients for wave modes. As we will show, for a massless
  conformal scalar field that represents scalar Hawking radiation, the
  radial wave equation can be transformed into the Heun equation,
  which has four regular singular points. In this case, the outer
  black hole horizon and the cosmological horizon correspond to
  regular singular points of the Heun equation.  Therefore, the
  asymptotic solutions at the horizon can be written with the local
  regular solutions of the Heun equation (local Heun function) via the
  Frobenius method, and it is possible to obtain the exact form of
  reflection and transmission coefficients in terms of the local Heun
  function. For the asymptotically flat Kerr spacetime, the radial
  wave function is represented by the confluent Heun function, which
  has two regular singular points and one irregular singular point,
  and spatial infinity corresponds to the irregular singular
  point. Although we have local solutions for the outer black hole
  horizon, it is technically difficult to match this solution to that
  of infinity. Concerning this issue, Hatsuda~\cite{Hatsuda2021}
  proposed a method of taking a small cosmological constant and
  extrapolating the value to obtain the quasi-normal frequency for the
  asymptotically flat black hole spacetime and further checked its
  validity.  In the present paper, we adopt his approach and
  investigate Hawking radiation in the KdS spacetime with a sufficiently
  small value of the cosmological constant, and the effect of radiation
  from the cosmological horizon is not significant.}  The vacuum
condition is imposed on the past event horizon and the past
cosmological horizon of the Kruskal extended KdS spacetime.  For
detecting Hawking radiation, we prepare two qubit detectors to
measure the spatial correlation of Hawking radiation.  Then, by the Fourier
transformation of the spatial correlation function, we can obtain wave
optical images of black holes.
  
%%%

The remainder of this paper is organized as follows. In Section II, we
shortly review the van Cittert-Zernike theorem and the qubit detector
model. We adopt two qubit detectors as our imaging system, which can
extract the spatial correlation of a wave field. In Section III, after
reviewing Hawking radiation in the KdS spacetime, we present the
spatial correlation function of Hawking radiation in terms of
transmission and reflection coefficients for wave modes. In Section
IV, we explain a numerical method to evaluate reflection and
transmission coefficients. In Section V, we present images of black
holes. Section VI is devoted to a summary and conclusion. We adopt
units of $c=\hbar=G=1$ throughout this paper.

\newpage
%%%%%%%%%%%%%%%%%%%%%%%%%%%%%%%%%%%%%%%%%%%%%%%%%%
\section{Van Cittert-Zernike theorem and wave optical imaging
}
\anno{ We shortly review a method of wave optical imaging based on the
  van Cittert-Zernike theorem~\cite{Born1999,Wolf2007,Sharma2006} for
  the flat spacetime,  which corresponds to asymptotically flat
  black hole spacetimes.  We also show in the Appendix that the same
  form of the theorem also holds for the de Sitter case by replacing the
  radial coordinate  in the phase factor with the tortoise coordinate
  of de Sitter space.
  Then, we review the qubit detector system, which is applicable to the
  detection of the  spatial correlations of Hawking radiation, to employ image
  formation based on the van Cittert-Zernike theorem.}

%%%%%%%%%%%%%%%%%%%%%%%%%%%%%%%%%%%%%55
\subsection{Van Cittert-Zernike theorem}
 
\begin{figure}[H]
  \centering
  \includegraphics[width=0.4\linewidth,clip]{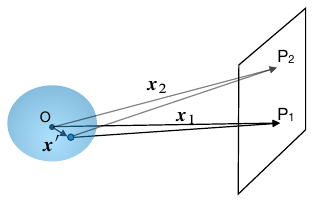}
  \caption{A wave source is located around the origin. Detection
    points $\textsf{P}_1$ and $\textsf{P}_2$ are assumed to be far from the source.}
  \label{fig:setup}
\end{figure}

Let us consider the emission of a scalar wave from a source
$\rho(t,\bs{x})$, which possesses a random statistical property. We
observe the emitted
wave far from the source (Fig.~\ref{fig:setup}).
%%%
%%%
\noindent 
We consider the scalar wave function $\Phi$ that obeys the  wave
equation $(\pa_t^2-\nabla^2)\Phi=\rho(t,\bs{x})$. \anno{Time
dependence is separated  using the  temporal Fourier components as
%%%
\begin{equation}
\Phi(t,\bs{x})=\int^{+\infty}_{-\infty}d\omega\,e^{-i\omega
  t}\Phi_\omega(\bs{x}),
\quad\rho(t,\bs{x})=\int^{+\infty}_{-\infty}d\omega\,e^{-i\omega t}\rho_\omega(\bs{x}),
\end{equation}
%%%
with $\Phi_\omega=\Phi^*_{-\omega}, \rho_\omega=\rho^*_{-\omega}$. The
wave equation is then expressed as
%%%
\begin{equation}
  (\nabla^2+\omega^2)\Phi_\omega=-\rho_\omega. \label{eq:fwave}
\end{equation}
%%%
The Green's function of this equation with a retarded boundary condition
is
%%%
\begin{equation}
  \mathcal{G}_\omega(\bs{x},\bs{x}_s)=\frac{e^{i\omega|\bs{x}-\bs{x}_s|}}{4\pi|\bs{x}-\bs{x}_s|}\approx\frac{e^{i\omega(r-\bs{x}\cdot\bs{x}_s/r)}}{4\pi
      r},
  \label{eq:Green}
\end{equation}
%%%%
where $r=|\bs{x}|$ and we assume the detection point $\bs{x}$ is far
from the source region. The solution of Eq.~\eqref{eq:fwave} is given
by
$\Phi_\omega(\bs{x})=\int d^3\bs{x}_s
\mathcal{G}_\omega(\bs{x},\bs{x}_s)\rho_\omega(\bs{x}_s)$. The
correlation function of the scalar field is
%%%
\begin{equation}
  \expval{\Phi(t_1,\bs{x}_1)\Phi(t_2,\bs{x}_2)}=\int_{-\infty}^{+\infty}d\omega\,e^{-i\omega(t_1-t_2)}\expval{\Phi_\omega(\bs{x}_1)\Phi^*_\omega(\bs{x}_2)},
\end{equation}
%%%
where  $\expval{\cdots}$ denotes statistical averaging and we used
the stationarity condition for the scalar field
$\expval{\Phi_{\omega_1}(\bs{x}_1)\Phi_{\omega_2}(\bs{x}_2)}\propto\del(\omega_1+\omega_2)$. The
correlation function of the temporal Fourier component of the field is
%%%
\begin{align}
 G(\omega,\bs{x}_1,\bs{x}_2):= \expval{\Phi_\omega(\bs{x}_1)\Phi^*_\omega(\bs{x}_2)}
  &=\int
    d^3\bs{x}_{s1}d^3\bs{x}_{s2}\mathcal{G}_\omega(\bs{x}_1,\bs{x}_{s1})\mathcal{G}^*_{\omega}(\bs{x}_2,\bs{x}_{s2})\expval{\rho_\omega(\bs{x}_{s1})\rho^*_{\omega}(\bs{x}_{s2})}\notag\\
  &=\int d^3\bs{x}_s\mathcal{G}_\omega(\bs{x}_1,\bs{x}_{s})\mathcal{G}^*(\bs{x}_2,\bs{x}_s)I_\omega(\bs{x}_s),
\end{align}
%%%
where we assumed spatial incoherency of the source field. This means
that the correlation between different spatial points is zero:
%%%
\begin{equation}
  \expval{\rho_\omega(\bs{x}_{s1})\rho^*_{\omega}(\bs{x}_{s2})}=I_\omega(\bs{x}_{s1})\del^3(\bs{x}_{s1}-\bs{x}_{s2}),
\end{equation}
%%%
where $I_\omega(\bs{x}_s)$ is the intensity of the source at $\bs{x}_s$.
Using \eqref{eq:Green} with $|\bs{x}_1|=|\bs{x}_2|=r$,
%%%
\begin{align}
  G(\omega,\bs{x}_1,\bs{x}_2)
  &=\frac{1}{16\pi^2r^2}\int
    d^3\bs{x}_s\,I_\omega(\bs{x}_s)\exp\left[-i\omega(\bs{x}_1-\bs{x}_2)\cdot\bs{x}_s/r\right]
    \notag \\
  &=\frac{1}{16\pi^2r^2}\int d^2\bs{x}_\parallel
    \widetilde I_\omega(\bs{x}_\parallel)
    e^{-i\omega\bs{x}_{12}\cdot\bs{x}_{\parallel}/r},
\end{align}
%%%
where we decompose $\bs{x}_s$ as
$\bs{x}_s=\bs{x}_\perp+\bs{x}_{\parallel},~
(\bs{x}_1-\bs{x}_2)\cdot\bs{x}_\perp=0$ and
$\bs{x}_{12}=\bs{x}_1-\bs{x}_2$.} The  projected source
intensity is introduced as
%%%
\begin{equation}
  \widetilde I_\omega(\bs{x}_\parallel)=\int dx_\perp\,
    I_\omega(x_\perp,
    \bs{x}_\parallel).
\end{equation}
%%%%
Ultimately, we obtain a relation between the spatial field correlation
function and the spatial distribution of source intensity:
%%%
\begin{align}
  G(\omega,\bs{x}_1,\bs{x}_2)
  &\propto\int d^2\bs{y}\,
    \widetilde I_\omega(\bs{y})\exp\left(-i\frac{\omega}{r}
    \bs{x}_{12}\cdot\bs{y}\right).
\end{align}
%%%%
This formula is called the van Cittert-Zernike
theorem~\cite{Born1999,Wolf2007,Sharma2006}.  Thus, we can reconstruct
the distribution of the source intensity (image of the source) from
the spatial field correlation function as follows:
%%%
\begin{equation}
  \widetilde I_\omega(\bs{y})\propto\int
  d^2\bs{x}_{12}\,G(\omega,\bs{x}_1,\bs{x}_2)
  \exp\left(i\frac{\omega}{r}\bs{y}\cdot\bs{x}_{12}\right).
  \label{eq:CZ2}
  \end{equation}
%%%
 Even if the property of spatial incoherence of
  the source is unknown, $\widetilde{I}_\omega(\bs{y})$ obtained using
  Eq.~\eqref{eq:CZ2}  provides one possible
  visualization of the source field, irrespective of the spatial
  incoherence of  the source field.

  \anno{For asymptotic de Sitter spacetimes, as we explain the
    Appendix, the same relation \eqref{eq:CZ2} holds on replacing the
    radial coordinate with the tortoise coordinate, provided that the 
    impact parameters of the involved wave modes are shorter than the
    Hubble horizon length.}

%%%%%%%%%%%%%%%%%%%%%%%%%%%%%%%%%%%%%%%%%%%%%%%%%%%%
\subsection{Qubit detector and response functions}
As a measurement apparatus of the spatial correlation of Hawking
radiation, we introduce two detectors interacting with  Hawking
radiation and obtain the field correlation through correlation between two
detectors. The detectors are assumed to have two internal levels
(qubit) with the energy gap $\omega_0>0$. The interaction Hamiltonian
between qubits and the quantum scalar field $\hat\Phi$ (Hawking
radiation) is assumed to be
%%%
\begin{equation}
  \hat H_\text{int}=g_1(t)(\sigma_1^{+}+\sigma_1^{-})\hat\Phi(\bs{x}_1(t))
  +g_2(t)(\sigma_2^{+}+\sigma_2^{-})\hat\Phi(\bs{x}_2(t)),
\end{equation}
%%%
where $\sigma^{+}_{1,2}$ and $\sigma^{-}_{1,2}$ are raising and
lowering operators, respectively, for the detector's state and
$g_{1,2}(t)$ are switching functions. The world lines of detectors are
denoted by $\bs{x}_{1,2}(t)$. This detector system setup is often
employed to investigate the entanglement harvesting of quantum
fields~\cite{Reznik2003,Pozas-Kerstjens2015,Henderson2017,Nambu2011}. In
our investigation to measure the spatial correlation of Hawking
radiation, two detectors are placed at the same radial coordinate far
from the black hole.
%%%%
For the initial ground state of the detectors, after interaction with the
scalar field, the detector state
becomes~\cite{Nambu2011,Matsumura2020a}
%%%
\begin{equation}
  \rho_{12}=
  \begin{bmatrix}
    X_4&0&0&X\\0&E_1&E_{12}&0\\ 0&E_{12}&E_2&0\\ X^*&0&0&1-E_1-E_2-X_4
  \end{bmatrix},
  \label{eq:rho-state}
\end{equation}
%%%
where
%%%
\begin{align}
  &X=-2\int_{-\infty}^{+\infty}dt_1\int_{-\infty}^{t_1}dt_2\, g_1
    g_2\,e^{i\omega_0(t_1+t_2)}\expval{\hat\Phi(t_1,\bs{x}_1)\hat\Phi(t_2,\bs{x}_2)},\\
  &E_{12}=\int_{-\infty}^{+\infty}dt_1\int_{-\infty}^{+\infty}dt_2\,
    g_1g_2\,e^{-i\omega_0(t_1-t_2)}
    \expval{\hat\Phi(t_1,\bs{x}_1)\hat\Phi(t_2,\bs{x}_2)},\\
    &E_1=E_{12}|_{2=1},\quad E_2=E_{12}|_{1=2}, \quad X_4=O(g^4).
\end{align}
%%%
The expectation values of field operators are taken with respect to
the assumed quantum state of the scalar field.  The component
$E_{1,2}(\omega_0)$ represents the amount of local quantum fluctuation measured
by detectors and shows the Planckian distribution for black hole
cases~\cite{Unruh1976,Birrell1984}. The component $X$ represents quantum
coherence between two detectors. The entanglement between
two detectors can be judged by the entanglement
negativity~\cite{Vidal2002a}, which is proportional to
$|X|-\sqrt{E_1E_2}$ in the present case. Positive values
of this quantity imply that two detectors are entangled and entanglement
of the quantum field is measured by the detectors. Entanglement harvesting
in black hole spacetimes has been investigated in several studies (BTZ
case~\cite{Henderson2017}, Schwarzschild case~\cite{Andhini2017}, and Kerr
case~\cite{Menezes2018}).  The two-point function (Wightman function)
is expressed as
%%%
\begin{equation}
  D^{+}(t_1,t_2,\bs{x}_1,\bs{x}_2):=
  \expval{\hat\Phi(t_1,\bs{x}_1)\hat\Phi(t_2,\bs{x}_2)}=\int_{-\infty}^{+\infty}d\omega\, e^{-i\omega(t_1-t_2)}G(\omega,\bs{x}_1,\bs{x}_2)
\end{equation}
%%%
because of the stationarity of the correlation
$D^{+}(t_1,t_2,\bs{x}_1,\bs{x}_2)=D^{+}(t_1-t_2,\bs{x}_1,\bs{x}_2)$. By
changing the integration variables to
$ \anno{x}=(t_1+t_2)/2$ and $\anno{y}=(t_1-t_2)/2$, and assuming constant
switching functions $g_1=g_2=g$, we
obtain\footnote{$$
  G(-\omega,\bs{x}_1,\bs{x}_2)=G^*(\omega,\bs{x}_1,\bs{x}_2),\quad
  G(-\omega,\bs{x},\bs{x})=G(\omega,\bs{x},\bs{x}).$$}
%%%
\begin{align}
  &X=-4g^2\int_{-\infty}^{+\infty}dx\,e^{2i\omega_0
    x}\int_0^{+\infty}dy\,D^{+}(2y,\bs{x}_1,\bs{x}_2)\propto
    g^2\del(\omega_0)
    ,\\
  &E_{12}=2g^2\int_{-\infty}^{+\infty}dx\int_{-\infty}^{+\infty}dy\,e^{-2i\omega_0
    y}D^{+}(2y,\bs{x}_1,\bs{x}_2)=4\pi g^2\left(\int
    dx\right)G({-\omega_0},\bs{x}_1,\bs{x}_2), \label{eq:E12}\\
  &E_1=\left.E_{12}\right|_{\bs{x}_2=\bs{x}_1},\quad
    E_2=\left.E_{12}\right|_{\bs{x}_1=\bs{x}_2}.
\end{align}
%%%
The formal expression \eqref{eq:E12} contains an infinite factor, but
  it should be treated with some cutoff of integration \annor{and $G$
    is replaced by the Fourier transformation with finite interval of
    the correlation function.}
\anno{Because $\omega_0\neq 0$, we have $X=0$ for  constant switching
  functions.}  \anno{ By considering the state tomography of the detector
  system, that is, by measuring components of the state
  \eqref{eq:rho-state}, it is possible to access the component
  $E_{12}$, which is proportional to the temporal Fourier component of
  the Wightman function $G(\omega_0,\bs{x}_1,\bs{x}_2)$, and this
  quantity represents the spatial correlation of the quantum field.
  Therefore, the setup of two detectors can be applied as an
  imaging system of black holes with Hawking radiation.}

%%%%%%%%%%%%%%%%%%%%%%%%%%%%%%%%%%%% 
\section{Hawking radiation in Kerr-de Sitter  spacetime}
We shortly review  Hawking radiation in the Kerr-de Sitter
spacetime~\cite{Unruh1976,Sataloff,Ottewill2000,Fro,Birrell1984}.

%%%%%%%%%%%%%%%%%
\subsection{Basic formulas}
\begin{figure}[H] 
  \centering 
  \includegraphics[width=0.95\linewidth,clip]{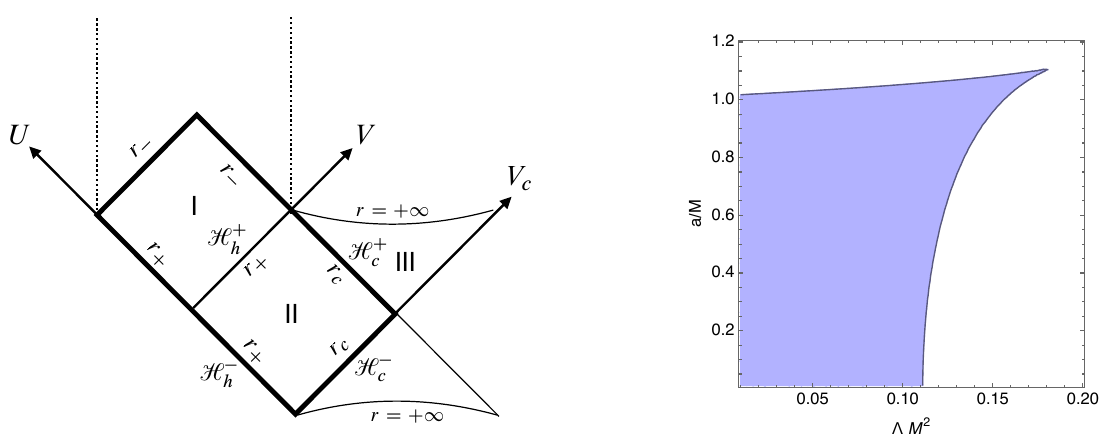}
  \caption{Left panel: global structure of the KdS spacetime considered to
    investigate Hawking radiation with the Unruh-Hawking vacuum state. Dotted
    vertical lines represent singularity. Right panel: colored region
    represents the parameters for a real $r_{+}$ and a real $r_c$. The upper
    boundary is the extremal limit, where $\kappa_{+}=0$. The lower
    boundary corresponds to $r_{+}=r_c$ (Nariai limit). Black holes
    with $a>1$ (``over spinning'' yet  maintaining its horizon
      structure) are possible for $\Lambda>0$. }
  \label{fig:penrose} 
\end{figure}

We consider a massless conformal scalar field $\varphi$ in the Kerr-de
Sitter (KdS) spacetime. This scalar field is equivalent to the scalar
mode of gravitational perturbation, which obeys the Teukolsky
equation. The metric of the KdS spacetime is
%%%
\begin{align}
  &ds^2=-\frac{\Delta_r}{\rho^2\chi^4}\left(dt-a\sin^2\theta
    d\phi\right)^2+\frac{\Delta_\theta}{\rho^2\chi^4}\sin^2\theta\left((r^2+a^2)
    d\phi-a
    dt\right)^2+\rho^2\left(\frac{dr^2}{\Delta_r}+\frac{d\theta^2}{\Delta_\theta}\right),
\end{align}
%%%
with
%%%
\begin{align}
  &\Delta_r=(r^2+a^2)\left(1-\frac{\Lambda}{3}r^2\right)-2Mr,\quad\Delta_\theta=
    1+\frac{\Lambda}{3}a^2\cos^2\theta,\\
  &\rho^2=r^2+a^2\cos^2\theta,\quad
  \chi^2=1+\frac{\Lambda}{3}a^2.
\end{align}
%%%
The parameters specifying this spacetime are $(M,a,\Lambda)$. 
The scalar field  obeys the following wave equation:
%%%
\begin{equation}
  -\nabla_\mu\nabla^\mu\varphi+\frac{1}{6} R^{(4)}\varphi=0,\quad
  R^{(4)}=4\Lambda, 
\end{equation}
%%%
where $R^{(4)}$ is the four-dimensional Ricci scalar.  We introduce the
tortoise coordinate $r_*$ defined by
%%%
\begin{equation}
  r_*=\int
  dr\frac{r^2+a^2}{\Delta_r}=\frac{\log|r-r_c|}{2\kappa_c}
  +\frac{\log|r-r_{+}|}{2\kappa_{+}} 
  +\frac{\log|r-r_{-}|}{2\kappa_{-}}
  +\frac{\log|r-r_{--}|}{2\kappa_{--}},
 \end{equation}
 %%%
 where $r_{--}<0<r_{-}<r_{+}<r_c$ are four roots of $\Delta_r=0$;
 $r_{+}$ is the radius of the outer event horizon, and $r_c$ is the
 radius of the cosmological horizon (Fig.~\ref{fig:penrose}).  The
 surface gravity $\kappa$ and the angular velocity $\Omega$ at these
 points are given by
 %%%
 \begin{equation}
   \kappa_j:=\kappa(r_j)=\frac{\Delta_r'(r_j)}{2\chi^2(r_j^2+a^2)},\quad
   \Omega_j=\frac{a}{r_j^2+a^2},\quad j=--,-,+,c,
 \end{equation}
 %%%
 \anno{where $'=\pa/\pa r$.}
 The right panel of Fig.~\ref{fig:penrose} shows a parameter region
 for a real $r_{+}$ and a real $r_c$ in the $(\Lambda M^2,a/M)$
 plane~\cite{Akcay2011}. For such values of parameters, we have a Kerr
 black hole enclosed by a cosmological horizon.
%%%
%%%  
%%%

The scalar field in the KdS spacetime is separated as
%%%
\begin{equation}
  \varphi_{\omega\ell m}=\frac{R_{\omega\ell
      m}(r)}{\sqrt{r^2+a^2}}S_{\omega\ell m}(\cos\theta)e^{im\phi},
  \label{eq:radial-func}
\end{equation}
%%%
where $S_{\omega\ell m}(\cos\theta)$ is the angular wave function obeying the
following equation:
%%% 
\begin{align}
  &\Biggl[\frac{d}{d\xi}\{1+(\chi^2-1)\xi^2\}(1-\xi^2)\frac{d}{d\xi}
    \notag\\
  &\qquad\qquad
    -2(\chi^2-1)\xi^2
    -\frac{\chi^4\{m-(1-\xi^2)a\omega\}^2}{\{1+(\chi^2-1)\xi^2\}(1-\xi^2)}
    +\lambda_{\ell
    m}(a\omega,\Lambda)\Biggr]S_{\omega\ell m}(\xi)=0,
    \label{eq:angular}
\end{align}
%%%
where $\xi=\cos\theta$ and $\lambda_{\ell m}(a\omega,\Lambda)$ is the
eigenvalue of this equation.  As Eq.~\eqref{eq:angular} has
four regular singular points, it can be written in terms of the Heun equation
with an appropriate transformation \cite{Hatsuda2021,Motohashi2021a}
to obtain the values of $S_{\omega\ell m}(\xi)$ for the range of
$0\leq \theta \leq \pi$ using the local Heun functions. Moreover, the
eigenvalues $\lambda_{\ell m}(a\omega,\Lambda)$ are also obtained by
finding the zero point of the Wronskian for the linear independent
local Heun functions, which is equivalent to the regularity condition
for $S_{\omega\ell m}(\xi)$ at $\theta=0,\pi$
\cite{Hatsuda2021,Motohashi2021a}.
%
% As shown in \anno{[Fig.??]}, the contribution of $\Lambda$ to $S_{\ell m}$ is
%very small even for $\Lambda=0.1$, therefore in the  
%present paper, we use the spheroidal harmonics without $\Lambda$,
%which is given as the built-in function {\tt SpheroidalPS} in
%Mathematica.
For the eigenvalues, an analytic formula was derived
in~\cite{1386}. \anno{ We have checked that the formula provides
  numerical values of $\lambda_{\ell m}$ with acceptable accuracy even
  for the parameter region $a\omega\sim O(1)$ with $\Lambda=1/100$. Hence
  we used the analytic formula in the present study.}

The radial wave function $R_{\omega\ell m}(r)$ obeys
%%%
\begin{equation}
  \left[\frac{d^2}{dr_*^2}-V_{\omega\ell m}(r)\right]R_{\omega\ell
    m}=0
  \label{eq:radial-eq}
\end{equation}
%%%
with the potential
%%%
\begin{align}
  V_{\omega\ell
  m}(r)&=-\chi^4\left(\omega-m\Omega\right)^2 \notag \\
 & +\frac{\Delta_r}{\chi^4(r^2+a^2)^2}\left\{
  \lambda_{\ell m}(a\omega,\Lambda)+\frac{2}{3}\Lambda
  r^2+(r^2+a^2)^{1/2}\left(\frac{r\Delta_r}{(r^2+a^2)^{3/2}}\right)'\right\}.
\end{align}
%%%%
For investigating  Hawking radiation, we consider
the extended KdS spacetime shown in the left panel of
Fig.~\ref{fig:penrose}.  The Kruskal coordinates about bifurcating
horizons $r_{+}$ and $r_c$ are defined as
%%%
\begin{align}
  &U=+\frac{e^{-\kappa_{+} u}}{\kappa_{+}},\quad
    V=\frac{e^{\kappa_{+} v}}{\kappa_{+}}\quad \text{in region I},\\
  &U=-\frac{e^{-\kappa_{+} u}}{\kappa_{+}},\quad
    V=\frac{e^{\kappa_{+} v}}{\kappa_{+}},\quad
    V_c=-\frac{e^{-\kappa_c v}}{\kappa_c}\quad\text{in 
    region II},\\
  &V_c=\frac{e^{-\kappa_c v}}{\kappa_c}\quad \text{in region III},
\end{align}
%%%
where $u=t-r_*$ and $v=t+r_*$.
%%%

As the state of Hawking radiation, we adopt the Unruh-Hawking vacuum
state. \anno{The condition for the quantum state is presented in the
  subsection B of the present section. We shortly comment on three
  typical vacuum states for quantum fields in the asymptotically flat
  static black hole spacetime. The Boulware vacuum state is defined by
  taking positive frequency modes for the past and future null
  infinities. This vacuum state shows no particle emission from a
  black hole. The Hartle-Hawking vacuum state is defined by taking
  incoming modes to be positive frequency modes with respect to
  $\pa_V$ ($V$ is the canonical affine parameter on the past black
  hole horizon) and outgoing modes to be positive frequency modes with
  respect to $\pa_U$ ($U$ is the canonical affine parameter on the
  future black hole horizon). This vacuum state is time symmetric and
  represents the thermal equilibrium state of a black hole and Hawking
  radiation. The Unruh-Hawking vacuum state is defined by taking modes
  incoming from the past null infinity to be positive frequency modes
  with respect to $\pa_t$ and those emanating from the past black hole
  horizon to be positive frequency modes with respect to $\pa_U$. This
  vacuum state is realized by the formation of a black hole via
  gravitational collapse. To specify the Unruh-Hawking vacuum state in
  the KdS spacetime, which is not asymptotically flat, we take modes
  incoming from the past cosmological horizon to be positive frequency
  modes with respect to $\pa_{V_c}$~\cite{Gregory2021}.}

\anno{To express the vacuum condition for the Unruh-Hawking vacuum, }
we first introduce the up mode and the dn mode; the mode
$\varphi^\text{up}$ has support only in  region II, and
$\varphi^\text{dn}$ has support only in  region I (see the left
panel of Fig.~2)\footnote{The modes of the wave equation are
  normalized with respect to the inner product
$$
  (\varphi^1,\varphi^2):=i\int_\Sigma d\sigma^\mu((\varphi^1)^*\pa_\mu\varphi^2-\varphi^2\pa_\mu(\varphi^1)^*),
  $$
  where $\Sigma$ is a spacelike or null hypersurface and $d\sigma^\mu$
  is the volume element on this surface.}; they are defined by
imposing their asymptotic forms at the past event horizon as
%%%
\begin{equation}
  \varphi^\text{up}|_{\mathcal{H}_h^{-}}\sim\exp\left(i\frac{\omega_{+}}{\kappa_{+}}\ln(-U)\right)\Theta(-U),\quad
  \varphi^\text{dn}|_{\mathcal{H}_h^{-}}\sim\exp\left(-i\frac{\omega_{+}}{\kappa_{+}}\ln(U)\right)\Theta(U),
\end{equation}
%%%
where \anno{$\omega_{+}=\omega-m\Omega_{+}>0$ and }$\Theta(x)$ is the
unit step function. \anno{Even for $\omega>0$, there is a possibility
  that $\omega_{+}<0$ and positive frequency modes with $\omega>0$
  become effectively negative frequency modes. These modes are called
  superradiant modes, which are peculiar to the Kerr spacetime.} Then, 
the UP mode, which is the outgoing positive frequency mode with respect to
the coordinate $U$ on the past event horizon $\mathcal{H}_h^{-}$
\anno{and analytic across the future event horizon
  $\mathcal{H}_h^{+}$}, is defined as a linear combination of
$\varphi^\text{up}$ and $\varphi^\text{dn}$ (for
$\omega,\omega_{+}>0$):
%%%
\begin{align}
  &\varphi^{\text{(UP1)}}_{\omega_{+}}=\frac{1}{\sqrt{2\sinh(\pi\omega_{+}/\kappa_{+})}}
    \left(
    e^{\pi\omega_{+}/2\kappa_{+}}\varphi^\text{up}_{\omega\ell
     m}+e^{-\pi\omega_{+}/2\kappa_{+}}(\varphi^\text{dn}_{-\omega\ell
     -m})^*\right),\\
  &\varphi^{\text{(UP2)}}_{-\omega_{+}}=
    \frac{1}{\sqrt{2\sinh(\pi\omega_{+}/\kappa_{+})}}
    \left(e^{\pi\omega_{+}/2\kappa_{+}}\varphi^\text{dn}_{-\omega\ell
   -m}+e^{-\pi\omega_{+}/2\kappa_{+}}(\varphi^\text{up}_{\omega\ell m})^*\right).
\end{align}
%%%
The IN mode, which is the incoming positive frequency mode with respect to
the coordinate $V_c$ on the past cosmological horizon
$\mathcal{H}_c^{-}$ \anno{and analytic across the future cosmological
  horizon $\mathcal{H}_c^{+}$}, is defined as
%%%
\begin{align}
  &\varphi^{\text{(IN1)}}_{\omega_c}=\frac{1}{\sqrt{2\sinh(\pi\omega_c/\kappa_c)}}
  \left(e^{\pi\omega_c/2\kappa_c}\varphi_\omega^\text{in}+e^{-\pi\omega_c/2\kappa_c}
    (\varphi_{-\omega}^\text{ot})^*\right),\\
  &\varphi_{-\omega_c}^{\text{(IN2)}}=\frac{1}{\sqrt{2\sinh(\pi\omega_c/\kappa_c)}}
    \left(e^{\pi\omega_c/2\kappa_c}\varphi_{-\omega}^\text{ot}
    +e^{-\pi\omega_c/2\kappa_c}
    (\varphi_{\omega}^\text{in})^*\right),
\end{align}
%%%%
with $\omega_c:=\omega-m\Omega_c$. The mode $\varphi^\text{in}_\omega$
has support only in the region II, $\varphi^\text{ot}_{-\omega}$ has
support only in  region III, and their asymptotic forms at the
  past cosmological horizon are specified by
%%%
\begin{equation}
  \varphi^\text{in}|_{\mathcal{H}_c^{-}}\sim\exp\left(-i\frac{\omega_{c}}{\kappa_{c}}\ln(-V_c)\right)\Theta(-V_c),\quad
  \varphi^\text{ot}|_{\mathcal{H}_c^{-}}\sim\exp\left(i\frac{\omega_{c}}{\kappa_{c}}\ln(V_c)\right)\Theta(V_c).
\end{equation}
%%%
  The asymptotic behaviors of radial functions
$R_{\omega\ell m}^\text{in}=\sqrt{r^2+a^2}\,\varphi_{\omega\ell
  m}^\text{in}$ and
$R_{\omega\ell m}^\text{up}=\sqrt{r^2+a^2}\,\varphi_{\omega\ell
  m}^\text{up}$ are
%%%
%%%
\begin{align}
  &R^\text{up}_{\omega\ell m}\rightarrow
    \begin{cases}
      e^{i\omega_{+} r_*}+\widetilde{\mathcal{R}}_{\omega\ell m}\,e^{-i\omega_{+}
        r_*},\quad &r_*\rightarrow-\infty ~(r\rightarrow r_{+})\\
      \widetilde{\mathcal{T}}_{\omega\ell m}\,e^{i\omega_c r_*},\quad
      &r_*\rightarrow+\infty~(r\rightarrow r_c) 
    \end{cases}, \label{eq:up-mode}\\
  &R^\text{in}_{\omega\ell m}\rightarrow
    \begin{cases}
      \mathcal{T}_{\omega\ell m}\,e^{-i\omega_{+} r_*},\quad &
      r_*\rightarrow-\infty~(r\rightarrow r_{+})\\
      e^{-i\omega_c r_*}+\mathcal{R}_{\omega\ell m}\,e^{i\omega_c
        r_*},\quad &r_*\rightarrow+\infty~(r\rightarrow r_c)
    \end{cases}, \label{eq:in-mode}
\end{align} 
%%%
where reflection coefficients $\mathcal{R}_{\omega\ell m}$ and
$\widetilde{\mathcal{R}}_{\omega\ell m}$, and transmission coefficients
$\mathcal{T}_{\omega\ell m}$ and $\widetilde{\mathcal{T}}_{\omega\ell m}$
are introduced. These coefficients satisfy the following relation,
which is originated from the conservation of the Wronskian:
%%%
\begin{align}
  &1-|\mathcal{R}_{\omega\ell
    m}|^2=\frac{\omega_{+}}{\omega_c}|\mathcal{T}_{\omega\ell
    m}|^2,\quad
    1-|\widetilde{\mathcal{R}}_{\omega\ell
    m}|^2=\frac{\omega_c}{\omega_{+}}|\widetilde{\mathcal{T}}_{\omega\ell
    m}|^2,   \label{eq:ref1}\\
  &\omega_c\,\widetilde{\mathcal{T}}^*_{\omega\ell m}\,\mathcal{R}_{\omega\ell
    m}=-\omega_{+}\,\mathcal{T}_{\omega\ell
    m}\,\widetilde{\mathcal{R}}^*_{\omega\ell m},\quad
    \omega_c\,\widetilde{\mathcal{T}}_{\omega\ell
    m}=\omega_{+}\,\mathcal{T}_{\omega\ell m}. \label{eq:ref2}
\end{align}
%%%

%%%%%%%%%%%%%%%%%%%%%%%%%%%%%%%%%%%%%%%%%%%%%%%%%%%%%%%%
\subsection{Correlation  function}
The introduced combinations of modes $\varphi^{(\text{UP})}$ and
$\varphi^{(\text{IN})}$ are called the Unruh modes. \annor{UP mode is
  defined on $\mathcal{H^{-}}$ and IN mode is defined on
  $\mathcal{I}^{-}$; Thus $\mathcal{H}^{-}\cup\mathcal{I}^{-}$ is the
  initial Cauchy surface to define the Unruh-Hawking vacuum state.}
Using the Unruh mode functions, the field operator is expanded as
%%%
\begin{align}
  \hat\Phi(x)&=\sum_{\ell m}\int_0^\infty d\omega_{+}\left(\hat
    a_{\omega_{+}}^\text{(UP1)}\varphi^\text{(UP1)}_{\omega_{+}}+\hat
    a_{-\omega_{+}}^\text{(UP2)}\varphi^\text{(UP2)}_{-\omega_{+}}\right)
               \notag\\
  &\qquad +\sum_{\ell
    m}\int_0^\infty d\omega_c
  \left(\hat a_{\omega_c}^\text{(IN1)}\varphi^\text{(IN1)}_{\omega_c}
    +\hat a^\text{(IN2)}_{-\omega_c}\varphi^\text{(IN2)}_{-\omega_c}
  \right)+\text{(h.c.)}.
\end{align}
%%%
The Unruh-Hawking vacuum state $\ket{U}$ 
%\footnote{\anno{We should add some comment on vacuum state in KdS
%  spacetime. difference of static case and stationary case }}
is defined by \anno{\cite{Gregory2021}}
%%%
\begin{equation}
  \hat a^\text{(UP1)}_{\omega_{+}}\ket{U}=\hat
  a^\text{(UP2)}_{-\omega_{+}}\ket{U}=\hat a_{\omega_c}^\text{(IN1)}\ket{U}=\hat a_{-\omega_c}^\text{(IN2)}\ket{U}=0,\quad
  \omega_{+},\omega_c\ge 0.
\end{equation}
%%%
This state is realized by black hole formation via gravitational
collapse in the KdS spacetime, and the UP  and  IN modes are
thermally populated at the \anno{past} black hole horizon and the
\anno{past} cosmological horizon, respectively. \anno{These modes are
  regular on $\mathcal{H}_h^{-}$ and $\mathcal{H}_c^{-}$.} The Hadamard's
elementary function with the Unruh-Hawking vacuum state is (we assume
$x_1,x_2\in\text{region II}$ in Fig.~\ref{fig:penrose})
\footnote{$(u_\omega(r))^*=u_{-\omega}(r)$.}
%%%
\begin{align}
  &G(x_1,x_2):=\langle
    U|\{\hat\Phi(x_1),\hat\Phi(x_2)\}|U\rangle
    \notag \\
  &=\sum_{\ell m}\int_0^\infty
    d\omega_{+}\coth\left(\frac{\pi\omega_{+}}{\kappa_{+}}\right)
    \varphi^\text{up}_{\omega\ell
    m}(x_1)\varphi^\text{up}_{\omega\ell
    m}{}^{\!\!\!\!*}(x_2)S_{\omega\ell m}(\xi_1)S_{\omega\ell m}(\xi_2)e^{im(\phi_1-\phi_2)} \notag \\
  &\quad +\sum_{\ell m}\int_0^\infty d\omega_c
    \coth\left(\frac{\pi\omega_c}{\kappa_c}\right)\varphi^\text{in}_{\omega\ell
    m}(x_1)\varphi^\text{in~*}_{\omega\ell
    m}(x_2)S_{\omega\ell m}(\xi_1)S_{\omega\ell m}(\xi_2)e^{im(\phi_1-\phi_2)}+(x_1\leftrightarrow
    x_2)+\text{(h.c.)}
    \notag \\
  &=\frac{1}{4\pi(r_1^2+a^2)^{1/2}(r_2^2+a^2)^{1/2}}\times \notag \\
  &\sum_{\ell
    m}\int_0^\infty d\omega\, e^{-i\omega(t_1-t_2)}\Biggl[\frac{\Theta(\omega_{+})}{\omega_{+}}\coth\left(\frac{\pi\omega_{+}}{\kappa_{+}}\right)R^{\text{up}}_{\omega\ell
    m}(r_1)R^{\text{up}~*}_{\omega\ell m}(r_2)
    \notag \\
  &+\frac{\Theta(\omega_c)}{\omega_c}
    \coth\left(\frac{\pi\omega_c}{\kappa_c}\right)
    R^{\text{in}}_{\omega\ell
    m}(r_1)R^{\text{in}~*}_{\omega\ell m}(r_2)\Biggr]S_{\omega\ell
    m}(\xi_1)S_{\omega\ell
    m}(\xi_2)e^{im(\phi_1-\phi_2)}+(1\leftrightarrow 2)+\text{(h.c.)},
\label{eq:Ga}
\end{align}
%%%%
where $\xi_{1,2}=\cos\theta_{1,2}$. To derive the last expression of
Eq.~\eqref{eq:Ga}, we have changed the integration variable from
$\omega_{+},\omega_c$ to $\omega$ and introduced the unit step
function in the integrand.  For $r_1=r_2\rightarrow r_c$, the temporal
Fourier component of the correlation function is given by
%%%
\begin{align}
  G(\omega,\bs{x}_1,\bs{x}_2)&\propto\sum_{\ell m}\Biggl[
                    \frac{\Theta(\omega_{+})}{\omega_{+}}
                    \coth\left(\frac{\pi\omega_{+}}{\kappa_{+}}\right)|
                    \widetilde{\mathcal{T}}_{\omega\ell 
                m}|^2\notag
  \\
              &\qquad
                +
                \frac{\Theta(\omega_c)}{\omega_c}\coth\left(\frac{\pi\omega_c}{\kappa_c}\right)
                |1+{\mathcal{R}}_{\omega\ell
                m}e^{i\del}|^2\Biggr]S_{\omega\ell
                m}(\xi_1)S_{\omega\ell
                m}(\xi_2)e^{im(\phi_1-\phi_2)}\notag \\
                             &\equiv G_1+G_2,
                             \label{eq:G}
\end{align}
%%%%
where a phase factor $\del=2\omega_c r_*$ is introduced.  As the
radial function $R^{\text{in}}$ is a liner combination of the incoming
wave and the reflected wave with amplitude $\mathcal{R}_{\omega\ell
    m}$, $G_2$ contains phase information  determined by the
  reflection coefficient and $\del$. Coefficients
$\coth(\pi\omega_{+}/\kappa_{+})$ and $\coth(\pi\omega_c/\kappa_c)$
reflect the thermal property of the black hole horizon and the
cosmological horizon, respectively.  For $\omega_+\rightarrow 0$,
$|\widetilde{\mathcal{T}}_{\omega\ell m}|\rightarrow \omega_+$ (see
Eq.~\eqref{eq:ref2}) and $G_1(\omega)$ is finite, whereas $G_2$
diverges for $\omega_c\rightarrow 0$.

The Fourier component of Hadamard's elementary function $G$ consists
of the contribution $G_1$ of the UP mode and $G_2$ of the IN
mode. $G_1$ represents the illumination of the black hole by both the
thermally populated UP mode with the Hawking temperature
$\kappa_{+}/(2\pi)$ and vacuum fluctuation from the inside of the
photon sphere. On the other hand, $G_2$ is the contribution of the IN
mode and represents scattering of the incoming thermal radiation with
temperature $\kappa_c/(2\pi)$ from the cosmological horizon and the
vacuum fluctuation by the black hole.  These types of radiations
illuminates the black hole from the outside of the photon sphere.
Concerning the superradiant phenomena, $G_1$ includes no superradiant
modes, because it only contains $\omega_{+}>0$ modes, whereas $G_2$
includes superradiant modes $\omega_{+}<0<\omega_c$ and can potentially
show the superradiant scattering effect.

In $G_2$, the phase factor originating from the reflection coefficient
$\mathcal{R}_{\omega\ell m}e^{i\del}$ provides the interference term
between incoming and reflected waves. As the behavior of the
interference term depends on $r_*$, and for the purpose of qualitative
understanding of images of the black hole, it is convenient to
evaluate $G_2$ by replacing
$|1+\mathcal{R}_{\omega\ell m}e^{i\del}|^2$ with
$1+|\mathcal{R}_{\omega\ell m}|^2$, which corresponds to the dropping
of the interference term between the incoming and the reflected
waves \anno{by hand}.  For this purpose, we introduce the correlation
function without the interference term as
%%%
\begin{equation}
  \widetilde G_2:=\sum_{\ell
    m}\frac{\Theta(\omega_c)}{\omega_c}\coth\left(\frac{\pi\omega_c}{\kappa_c}
  \right)
  \left(2-\frac{\omega_c}{\omega_{+}}|\widetilde
      {\mathcal{T}}_{\omega\ell m}|^2\right)S_{\omega\ell m}(\xi_1)S_{\omega\ell
      m}(\xi_2)e^{im(\phi_1-\phi_2)},
\end{equation}
%%%
where we used the relations \eqref{eq:ref1} and \eqref{eq:ref2} to
express $|\mathcal{R}_{\omega\ell m}|^2$ using
$|\widetilde{\mathcal{T}}_{\omega\ell m}|^2$. The correlation function
$G=G_1+\widetilde G_2$ neglects the interference term between incoming
and reflected waves. \annor{In the eikonal limit, as wave optical
  images obtained using $\widetilde G_2$ correspond to images obtained
  by the ray tracing method in geometric optics, it is possible to
  identify wave effect in images by comparing images with $G_2$ and
  those with $\widetilde G_2$.}

To extract the pure thermal effect of Hawking radiation, we express
$G(\omega)$ for the Boulware vacuum, which includes no thermal emission
from the black hole horizon and the cosmological horizons. The form of
the correlation function for this vacuum state formally obtained by
taking the limits of $\kappa_{+}\rightarrow 0$ and $\kappa_c\rightarrow 0$ in
\eqref{eq:G}:
%%%
\begin{align}
  &G^\text{Boulware}(\omega)\notag \\
  &\propto\sum_{\ell m}
    \left[\frac{\Theta(\omega_{+})}{\omega_{+}}
      |\widetilde{\mathcal{T}}_{\omega\ell m
          }|^2
          +\frac{\Theta(\omega_c)}{\omega_c}|1+\mathcal{R}_{\omega\ell m}
          e^{i\del}|^2\right]S_{\omega\ell m}(\xi_1)S_{\omega\ell m}(\xi_2)
                               e^{im(\phi_1-\phi_2)} \notag\\
  &=\sum_{\ell m}
    \left[\frac{\Theta(\omega_{+})}{\omega_c}
    (2+2\mathrm{Re}[\mathcal{R}_{\omega\ell m}
    e^{i\del}])
    +\frac{\Theta(-\omega_{+})}{\omega_c}|1+\mathcal{R}_{\omega\ell
    m}  e^{i\del}|^2\right]S_{\omega\ell m}(\xi_1)S_{\omega\ell m}(\xi_2)  e^{im(\phi_1-\phi_2)},
\end{align}
%%%
where we assume that there are no superradiant modes associated with the
cosmological horizon ($\omega_c>0$).  The contribution of particle
creations from the black hole and the cosmological horizon in the
correlation  function is encoded in the following  two point
functions obtained by subtracting the contribution of the Boulware
  vacuum, and this correlation function includes the Planckian factor:
%%%
\begin{align}
    &G(\omega)-G^\text{Boulware}(\omega)\propto \notag \\
    &\qquad\sum_{\ell m}
      \left[\frac{\Theta(\omega_{+})}
      {e^{2\pi\omega_{+}/\kappa_{+}}-1}
       \frac{|\widetilde{\mathcal{T}}_{\omega\ell m}|^2}{\omega_{+}}
      +\frac{\Theta(\omega_c)}{e^{2\pi\omega_c/\kappa_c}-1}
      \frac{|1+\mathcal{R}_{\omega\ell m}e^{i\del
      }|^2}{\omega_c}\right]S_{\omega\ell m}(\xi_1)S_{\omega\ell m}(\xi_2)e^{im(\phi_1-\phi_2)}.
      \label{eq:G-planck}
\end{align}
%%%
By definition, $G-G^{\text{Boulware}}$ becomes zero for
$\kappa_{+},\kappa_c\rightarrow 0$. We use this correlation function
for images of the black hole that directly reflect the Hawking effect.
% \footnote{Subtraction of
%  the contribution from the Boulware vacuum is equivalent to assuming
%  the rotation wave approximation for the interaction Hamiltonian
%  between the quantum field and the qubit. Under this assumption, the
%  qubit detector has no response to the vacuum fluctuation.}

\newpage
%%%%%%%%%%%%%%%%%%%%%%%%%%%%%%%%%%%%
\section{Evaluation of transmission and reflection coefficients}
In this section, we introduce our computation of the greybody factor
$|\widetilde{\mathcal{T}}_{\omega\ell m}|^2$ and the reflection
coefficient $\mathcal{R}_{\omega\ell m}$.  We adopt the method
developed in~\cite{Motohashi2021a}, which utilizes the local solutions
around regular singular points of the Heun equation (local Heun
function) with the Frobenius method to construct solutions of the
Teukolsky equation. The local Heun functions have been implemented as
built-in functions in Mathematica version 12.1 released in 2020.
%The greybody factor $|\tilde{\mathcal{T}}_{\omega\ell m}|^2$ and the
%reflection coefficient $\mathcal{R}_{\omega\ell m}$ are obtained by
%numerically. We adopt the Mathematica package used in~\cite{Motohashi2021} which can evaluate exact values of the
%greybody factor and the reflection coefficient utilizing \anno{the local solutions around regular singular points} of the Heun equation implemented in the Mathematica.
Here, we discuss the relation between scattering problems based on the
Teukolsky radial function $R^{(T)}$ and on the radial function $R$
introduced in \eqref{eq:radial-func}.  They are related as
$R^{(T)}=R/\sqrt{r^2+a^2}$. $R^{(T)}$ obeys
%%%
\begin{equation}
  \left[\frac{d}{dr}\Delta_r\frac{d}{dr}+\frac{\chi^4}{\Delta_r}\left[\omega(r^2+a^2)-am\right]^2-\frac{2\Lambda}{3}r^2-\lambda_{\ell m}(a\omega,\Lambda)\right]R^{(T)}=0.
\end{equation}
%%%
This equation is the same as that of the massless conformal scalar field
in the KdS spacetime.  To use the method with the local Heun function,
we transform the above equation into the Heun equation by introducing
the coordinate transformation from $r$ to $z$ and the redefinition of
the radial equation:
\begin{equation}
       z=\frac{r_c - r_-}{r_c - r_+} \frac{r-r_+}{r-r_-},
       \quad
    R^{(T)}=z^{B_1}(z-1)^{B_2}(z-z_r)^{B_3}(z-z_\infty)y^{(\rm r)}(z),
\end{equation}
and the radial equation yields
%%%
\begin{equation} \label{eq:yradTeu} \frac{d^2y^{\rm (r)}}{dz^2} + \left( \frac{2B_1+1}{z} + \frac{2B_2+1}{z-1} + \frac{2B_3+1}{z-z_r} \right) \frac{dy^{\rm (r)}}{dz} + \frac{(1-2B_4) z+v}{z(z-1)(z-z_r)}y^{\rm (r)} = 0, 
\end{equation}
%%%
where
%%%
\begin{equation}
z_\infty=\frac{r_c - r_-}{r_c - r_+} ,\quad z_r=z_\infty \frac{r_{--}
  - r_+}{r_{--} - r_-} ,\quad
B_j=i\frac{\chi^2(r_j^2+a^2)(\omega-m\Omega_j)}{\Delta_r'(r_j)},
\end{equation}
%%%
with $B_1=B_{+}, B_2=B_{c}, B_3=B_{--}, B_4=B_{-}$, and
%%%
\begin{equation}
v= \frac{ \lambda_{\ell m}  - (\Lambda/3) (r_+ r_- + r_c r_{--}) }{(\Lambda/3) (r_- - r_{--}) (r_+ - r_c) } 
- \frac{ i  [ 2 \chi^2 \{ \omega (r_+ r_- + a^2) - a m \}  ] }{(\Lambda/3) (r_- - r_{--}) (r_- - r_+) (r_+ - r_c)} .
\end{equation}
%%%
The sets of the linear independent local solutions of
Eq.~\eqref{eq:yradTeu} at $z=0$ (black hole outer horizon) and at
$z=1$ (cosmological horizon) are represented as ($y_{01}, y_{02})$ and
($y_{11}, y_{12})$, respectively \cite{Motohashi2021a}.  These
solutions are related to each other as
\begin{align}
y_{01}(z)&= C_{11}\,y_{11}(z)+C_{12}\,y_{12}(z) ,\quad y_{02}(z)= C_{21}\,y_{11}(z)+C_{22}\,y_{12}(z), \\
y_{11}(z)&= D_{11}\,y_{01}(z)+D_{12}\,y_{02}(z)
 , \quad y_{12}(z) = D_{21}\,y_{01}(z)+D_{22}\,y_{02}(z),
\end{align}
with the connection coefficients
%%%
\begin{equation} \label{C11} C_{11}=\frac{W_z[y_{01}, y_{12}]}{W_z[y_{11}, y_{12}]}, \quad
C_{12}=\frac{W_z[y_{01}, y_{11}]}{W_z[y_{12}, y_{11}]}, \quad 
C_{21}=\frac{W_z[y_{02}, y_{12}]}{W_z[y_{11}, y_{12}]}, \quad 
C_{22}=\frac{W_z[y_{02}, y_{11}]}{W_z[y_{12}, y_{11}]},
\end{equation}
%%%
and
\begin{equation} \label{D11} D_{11}=\frac{W_z[y_{11},y_{02}]}{W_z[y_{01},y_{02}]}, \quad
D_{12}=\frac{W_z[y_{11},y_{01}]}{W_z[y_{02},y_{01}]}, \quad 
D_{21}=\frac{W_z[y_{12},y_{02}]}{W_z[y_{01},y_{02}]}, \quad
D_{22}=\frac{W_z[y_{12},y_{01}]}{W_z[y_{02},y_{01}]},
\end{equation}
%%%
where $W_z[u,v]=u\, (dv/dz)-v\,(du/dz)$.  Note that the local
solutions are evaluated with the built-in function {\tt HeunG} in {\tt
  Mathematica}\footnote{See the detailed computation in
  \cite{Motohashi2021a}.}.

To obtain the greybody factor and reflection coefficient, we
investigate the behavior of the in and up modes in terms of $R(r_*)$
and $R^{T}(r)$.  In the tortoise coordinate $r_*$, the up and in modes
have been obtained as Eqs.~\eqref{eq:up-mode} and \eqref{eq:in-mode},
respectively. Equivalently, in the $r$ coordinate, it is expressed as
%%%
\begin{align}
  &R^{(T)\text{up}}\rightarrow
    \begin{cases}
 D^\text{(up)}\Delta_r^{B_1}+D^\text{(ref)}\Delta_r^{-B_1},\quad &
 r\rightarrow r_{+} \\
 D^\text{(trans)}\Delta_r^{B_2},\quad &r\rightarrow r_c
\end{cases}  \\
&R^{(T)\text{in}}\rightarrow
    \begin{cases}
 C^\text{(trans)}\Delta_r^{-B_1}, \quad &
 r\rightarrow r_{+} \\
 C^\text{(ref)}\Delta_r^{B_2}+C^\text{(inc)}\Delta_r^{-B_2},
 \quad &r\rightarrow r_c.
\end{cases}
\end{align}
%%%
The reflection coefficient ${{\mathcal{R}}}_{\omega \ell m}$ 
and the greybody factor $|\widetilde{{\mathcal{T}}}_{\omega \ell m}|^2$ 
can be written with the coefficients of the above solutions as
%%%
\begin{equation}
  {\mathcal{R}}_{\omega \ell m}=\frac{C^\text{(ref)}}{C^\text{(inc)}}
  ,\quad
  |\widetilde{\mathcal{T}}_{\omega \ell m}|^2=\left(\frac{r_c^2+a^2}{r_{+}^2+a^2}\right)\left|\frac{D^\text{(trans)}}{D^\text{(up)}}\right|^2
\end{equation}
%%%
by comparing the asymptotic form of $R(r_*)$ and $R^{(T)}$ using the relation between $r_*$ and $r$.
%onsidering the relationusing the conservation of the Wronskian for %%%
%$R_{\omega \ell m}(r_*)$ and $R^{(T)}(r)$:
%%%
%\begin{align}
%  &R_1\frac{d}{dr_*}R_2-R_2\frac{d}{dr_*}R_1=\text{const.%}, \\
%  &\Delta_r \left(R_1^{(T)}\frac{d}{dr}R_2^{(T)}-R_2^{(T)}\frac{d}{dr}R_1^{(T)}\right)=\text{const.}
%\end{align}
%%%
Furthermore, $C^{\text{(ref/up/trans)}}$ and
$D^{\text{(ref/up/trans)}}$ are represented with the connection
coefficients of the local Heun functions as demonstrated in
\cite{Motohashi2021a} as
%%%
\begin{align} 
C^{\rm (inc)} &= C_{22} (- 1)^{B_2} (1 - z_r)^{B_3} (1-z_\infty)
                (A_2)^{-B_2}, \label{eq:Cinc} \\
C^{\rm (ref)} &= C_{21} D^{\rm (trans)} \notag\\
              &= C_{21} (- 1)^{B_2} (1 - z_r)^{B_3}
                (1-z_\infty)(A_2)^{B_2} , \label{eq:Cref} \\
D^{\rm (up)} &= D_{11} (- 1)^{B_2} (- z_r)^{B_3} (- z_\infty) (A_1)^{B_1}, \\
  D^{\rm (ref)} &= D_{12} C^{\rm (trans)} \notag\\
&= D_{12} (- 1)^{B_2} (- z_r)^{B_3} (- z_\infty) (A_1)^{-B_1}, \label{eq:Ctrans} 
\end{align}
%%%
with
\begin{equation}
    A_1=\frac{z_\infty}{(r_+ -r_-)\Delta'(r_+)},
    \quad A_2=\frac{z_\infty (r_+-r_-)}{-(r_c-r_-)^2\Delta'(r_c)}.
\end{equation}
%%%

\newpage
%%%%%%%%%%%%%%%%%%%%%%%%%%%%%%%%%%%%%%%%%%%%%%%%%%%%%%%%%%%%%
\section{Imaging of black holes with Hawking radiation}
As we have shown in Eq.~\eqref{eq:E12}, the qubit detector system is
applicable to the detection of the spatial correlation of Hawking
radiation, and we adopt it as our imaging system for black holes.

%%%%%%%%%%%%%%%%%%%%%%%%%%%%%%%%%%%%%%%%%%%%%%%%%%%%%%%

%%%%%%%%%%%%%%%%%%%%%%%%%%%%%%
\subsection{Detail of imaging method and an example with a
    simple model}\label{sec:simple-model}
  In our imaging setup, two detectors are placed near the cosmological
  horizon. In the spherical coordinate system, detector 1 is placed at
  $(r,\theta_1,\phi_1)=(r,\pi/2,0)$ (on the equatorial plane), and
  detector 2 is placed at $(r,\theta_2,\phi_2)$.
%%%
In the Cartesian coordinates
$(x,y,z)=(r\sin\theta\cos\phi, r\sin\theta\sin\phi, r\cos\theta)$,
the detector locations are
%%%
\begin{equation}
  \bs{x}_1=(r,
    0,0),\quad\bs{x}_2=(r\sin\theta\cos\phi,
    r\sin\theta\sin\phi, r\cos\theta).
\end{equation}
%%%
We define two-dimensional coordinates in the observer's screen as
%%%
%%%
\begin{equation}
 \bs{X}= (X,Y)=(\sin\theta\sin\phi,\cos\theta).
\end{equation}
%%%
Then, the locations of two detectors in the observer's screen are
%%%
\begin{equation}
    (X_1,Y_1)=(0,0)=\bs{X}_1,
    \quad (X_2,Y_2)=(\sin\theta\sin\phi,\cos\theta)=\bs{X}_{2}.
\end{equation}
%%%
Applying Eq.~\eqref{eq:CZ2}, images (intensity distributions) are
obtained through the following Fourier transformation of the correlation
function in the observer's screen:
%%%
\begin{equation}
  \mathcal{F}[G](\bs{X}_\text{im}):=\int d^2\bs{X}_{12}\,
  G(\omega,\bs{X}_1,\bs{X}_2)\exp\left(i\omega
    \bs{X}_\text{im}\cdot\bs{X}_{12}
  \right),\quad\bs{X}_{12}=\bs{X}_1-\bs{X}_2,
  \label{eq:fourier}
\end{equation}
%%%
where $\bs{X}_\text{im}$ denotes coordinates in the image plane.

To check our imaging method, we analytically evaluate images
for the model correlation function $G=G_1+\widetilde G_2$, which
includes important features of the correlation function of
Hawking radiation \anno{from the Schwarzschild-de Sitter black hole:}
%%%
\begin{equation}
  G_1=\frac{g_1}{\omega}\sum_{\ell=0}^{\infty}\Theta(\ell_*-\ell)(2\ell+1)P_\ell(\cos
  \Delta),\quad
  \widetilde
  G_2=\frac{g_2}{\omega}\sum_{\ell=0}^\infty\left(1+\Theta(\ell-\ell_*)\right)(2\ell+1)P_\ell(\cos \Delta).
  \label{eq:model-G}
\end{equation}
%%%%
\anno{Here, $P_\ell$ is the Legendre polynomial.} The transmission and
reflection coefficients are replaced with the unit step function to
reflect the property of perfect absorption of black holes. The
parameter $\ell_*$ denotes the critical angular momentum of perfect
absorption corresponding to the photon sphere, and
$\Delta:=\theta-\pi/2$ is the angle between an observing point and the
optical axis. \anno{In the eikonal limit, the critical impact
  parameter of photons is $\ell_*/\omega$ which corresponds to the size of
  the black hole shadow.} This model well represents the correlation
function of Hawking radiation in the eikonal region. The constants
$g_1$ and $g_2$ depend on the surface gravity of horizons
%%%
\begin{equation}
  g_1=\coth\left(\frac{\pi\omega}{\kappa_{+}}\right),\quad
  g_2=\coth\left(\frac{\pi\omega}{\kappa_{c}}\right).
\end{equation}
%%%
By replacing the sum with an integral, it is possible to evaluate the
correlation function analytically; with the approximation
$P_\ell(\cos\Delta)\sim J_0(\ell\Delta),~\ell\gg 1$ and by replacing
the upper bound of the infinite sum by $\ell_\text{max}\gg\ell_*$, we
obtain
%%%
\begin{align}
  G_1&\approx\frac{2g_1}{\omega}\int_0^{\ell_*}d\lambda\, \lambda
       P_\lambda(\cos\Delta)
  =\frac{2g_1}{\omega}\int_0^{\ell_*}d\lambda\,\lambda
    J_0(\lambda\Delta)=
    \frac{2g_1\ell_*}{\omega\Delta}J_1(\ell_*\Delta),
\end{align}
%%%
\anno{where $J_0$ and $J_1$ are Bessel functions} and
%%%
\begin{align}
  \widetilde G_2&=
       \frac{g_2}{\omega}\sum_{\ell=0}^{\ell_\text{max}}(2\ell+1)P_\ell(\cos\Delta)+
               \frac{g_2}{\omega}\left(\sum_{\ell=0}^{\ell_\text{max}}
               -\sum_{\ell=0}^{\ell_*}
               \right)
       (2\ell+1)P_\ell(\cos\Delta)\notag
  \\
    &\approx\frac{4g_2}{\omega}\frac{\ell_\text{max}}{\Delta}J_1(\ell_\text{max}\Delta) 
               -\frac{2g_2}{\omega}
    \frac{\ell_*}{\Delta}J_1(\ell_*\Delta).
\end{align}
%%%
%%%%%%%%%%
The angle $\Delta$ is related to the coordinates $(X,Y)$ in the observer's
screen as $\anno{|\Delta|}=\sqrt{X^2+Y^2}<1$.
%%%
%%%
\begin{figure}[H]
  \centering
  \includegraphics[width=0.9\linewidth,clip]{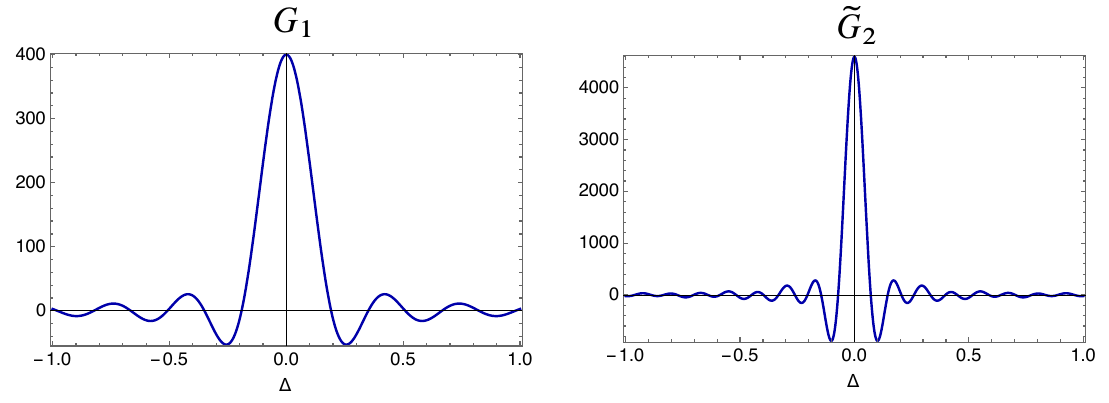}
  \caption{$G_1$ and $\widetilde{G}_2$ with $\ell_*=20, \ell_{\text{max}}=50,
    \omega=1$, and $g_1=g_2=1$.}
  \label{fig:G12}
\end{figure}
%%%
\noindent
Figure~\ref{fig:G12} shows the behavior of $G_1$ and $\widetilde G_2$. $G_1$
represents the interference fringe due to  Hawking radiation from the
black hole. For $\ell_\text{max}\rightarrow\infty$, the peak of
$\widetilde G_2$ at $\Delta=0$ becomes infinite, and the  peak approaches
the Dirac delta function.

\begin{figure}[H]
  \centering
  \includegraphics[width=0.9\linewidth,clip]{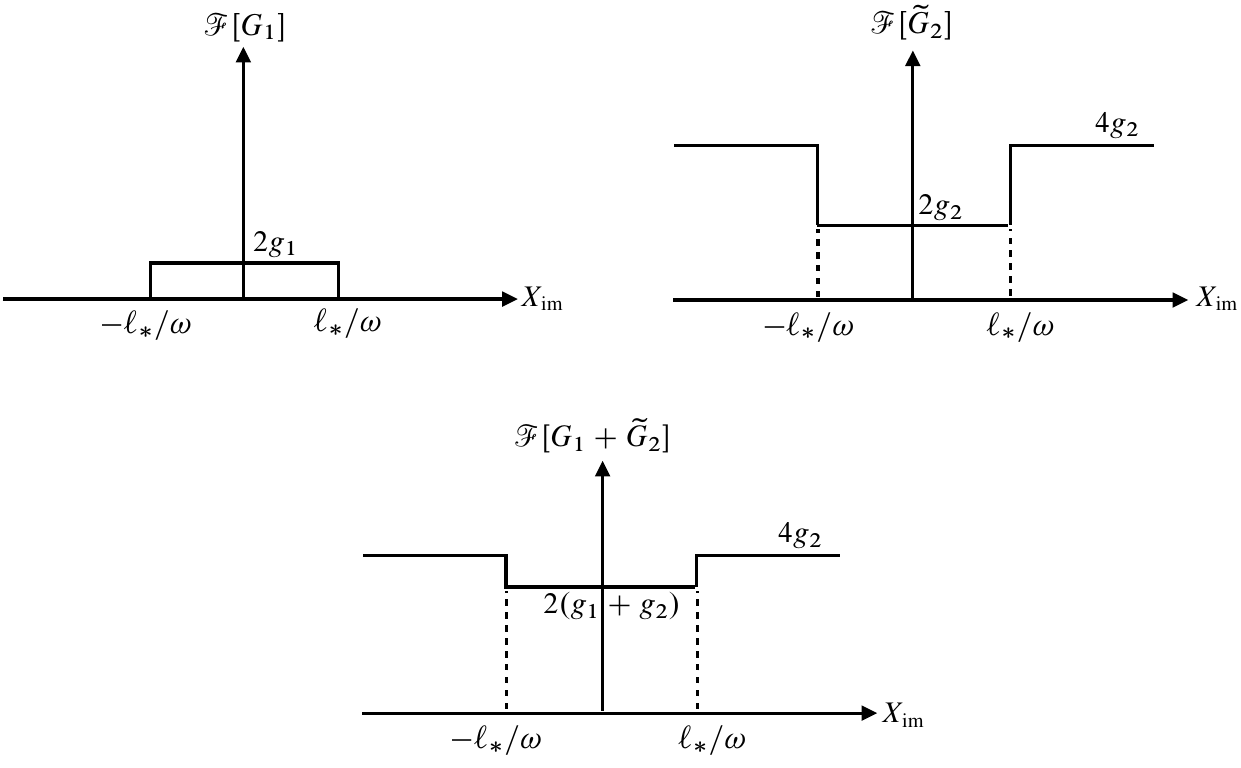}
  \caption{One-dimensional images obtained from the correlation
    function \eqref{eq:model-G}. The lower panel is
    $\mathcal{F}[G_1+\widetilde{G}_2]$. \anno{The red line is for the
      case of $g_1<g_2 ~(\kappa_{+}<\kappa_c)$ and the blue line is
      for the case of $g_1>g_2 ~(\kappa_{+}>\kappa_c)$. For high
      frequency, $g_1=g_2$, and $\mathcal{F}[G_1+\widetilde G_2]$
      is constant.}}
  \label{fig:toy}
\end{figure}

Now, we consider the two-dimensional Fourier transformation
\eqref{eq:fourier} of a function $f(\sqrt{X^2+Y^2})$:
%%%
\begin{equation}
  \mathcal{F}[f]=\frac{1}{2\pi}\int_{-\infty}^{+\infty} dX dY
  f(\sqrt{X^2+Y^2})\,e^{i\omega(X_\text{im}X+Y_\text{im}Y)},
\end{equation}
%%%
where $X_\text{im}$ and $Y_\text{m}$ are coordinates in the image
plane. Then, 
%%%
\begin{align}
  \mathcal{F}[f]
  &=\frac{1}{2\pi}\int_0^\infty \Delta (d\Delta)\int_0^{2\pi}d\phi
    f(\Delta)e^{i\omega\sqrt{X_\text{im}^2+Y_\text{im}^2}\,\Delta\cos\phi} \notag\\
 &=\int_0^\infty \Delta (d\Delta)
   J_0\left(\omega R_\text{im}\Delta\right)
   f(\Delta),
\end{align}
%%%
where $R_\text{im}:=\sqrt{X_\text{im}^2+Y_\text{im}^2}$. Applying this
formula\footnote{$$ \int_0^\infty dx J_0(a x)J_1(b
  x)=\frac{1}{b}\Theta(b-a).$$}, we obtain
%%%
\begin{align}
  \mathcal{F}[G_1]&\propto\frac{2g_1}{\omega}\Theta(\ell_*/\omega-R_\text{im}), \\
  \mathcal{F}[\widetilde G_2]&\propto \frac{4g_2}{\omega}\Theta(\ell_\text{max}/\omega-R_\text{im})-\frac{2g_2}{\omega}\Theta(\ell_*/\omega-R_\text{im}), \\
  \mathcal{F}[G_1+\widetilde G_2]&\propto \frac{4g_2}{\omega}\Theta(\ell_\text{max}/\omega-R_\text{im})
                        -\frac{2g_2-2g_1}{\omega}\Theta(\ell_*/\omega-R_\text{im}).
\end{align}
%%%
Figure~\ref{fig:toy} shows images obtained from the correlation
function \eqref{eq:model-G}. 
%%%
%%%
\noindent 
For high frequency,  $\omega>\kappa_{+,c}/(2\pi)$, and
$g_1\approx g_2\approx 1$ holds. \anno{The image $\mathcal{F}[G_1]$ is
  a bright disk with the intensity $2g_1$, which represents Hawking radiation
  from the black hole. On the other hand, the image
  $\mathcal{F}[\widetilde G_2]$ shows a dark shadow of the black hole
  in bright background, which is originated from emission of the
  cosmological horizon.} $\mathcal{F}[G_1+\widetilde G_2]$ is
constant, and the images $\mathcal{F}[G_1]$ and
$\mathcal{F}[\widetilde G_2]$  complement each other. On the other
hand, for low frequency  $\omega<\kappa_{+,c}/(2\pi)$,
$g_1\approx \kappa_{+}/(2\pi\omega)$, and $g_2\approx
\kappa_c/(2\pi\omega)$. Thus, for $\kappa_{+}<\kappa_c$, emission from
the cosmological horizon has a higher temperature than that of the black
hole, and the image $\mathcal{F}[G_1+\widetilde G_2]$ shows a dark
shadow with radius $\ell_*/\omega$.  On the contrary, for
$\kappa_{+}>\kappa_c$, the image $\mathcal{F}[G_1+\widetilde G_2]$
shows a bright disk, which represents the emission of Hawking radiation
from the black hole.

%%%%%%%%%%%%%%%%%%%%
\subsection{Black hole images} {Now, we proceed to image
  reconstruction of black holes using Eqs.~\eqref{eq:G} and
  \eqref{eq:G-planck}.  Assuming that detector 1 is located on the
  equatorial plane reduces much computational time because in the
  summation with respect to $\ell$ and $m$ of $G_1$ and $G_2$, the
  spheroidal harmonics $S_{\omega\ell m}(\pi/2,0)$ is nonzero only for
  $m=\ell,\ell-2,\cdots,-(\ell-2),-\ell$.  To evaluate $G_2$, we must
  truncate infinite sum of $\ell$ with a sufficiently large value
  $\ell_\text{max}$ that does not change the qualitative behavior of
  the correlation function. A rough estimation to determine
  $\ell_\text{max}$ is as follows: for the radial distance of
  observation $r_\text{obs}$, $\ell_\text{max}$ is estimated as
  $r_\text{obs}\sim \ell_\text{max}/\omega$.  Thus,
  $\ell_\text{max}\gtrsim\omega r_\text{obs}$ is required.  In our
  calculation, we choose $\ell_\text{max}=70$ for $\omega=5$ and
  $\ell_\text{max}=7$ for $\omega=0.5$. These values are chosen to be
  larger than the value of $\ell$ corresponding to the photon sphere
  of the black hole. The original correlation function,
  Eq.~\eqref{eq:Ga}, includes the prefactor
  $1/(4\pi(r_{\text{obs}}^2+a^2))$, which depends on
  $r_\text{obs}$. This factor only affects the total intensity of
  images, and the structure of images is not altered if we omit this
  factor. $G_2$ contains $r_{\text{obs}}$ as the phase factor
  $\del=2\omega_c r_*|_{\text{obs}}$. In our analysis, we do not fix
  $r_\text{obs}$, and $\delta$ is chosen as $0,\pi/2,\pi, 3\pi/2$.  As
  the black hole parameters, we choose $a=0, ~0.1$, and $1$ as well as 
  $\Lambda=1/100$. For these values, the horizon radius and surface
  gravity are (in units of $M=1$)
%%%
\begin{align}
        &a=0:  &r_{+}=2.028, \quad &r_c=16.22, \quad\kappa_{+}=0.2364,
                                                 \quad \kappa_c=0.05026
          ,\notag \\
        &a=0.1:  &r_{+}=2.023, \quad &r_c=16.22, \quad\kappa_{+}=0.2359,
                                                   \quad\kappa_c=0.05025, \\
        &a=1:  &r_{+}=1.094, \quad &r_c=16.22, \quad\kappa_{+}=0.03685,
                                                 \quad\kappa_c=0.05015. \notag
\end{align}
%%% 

In the observer's screen, we evaluate $G_1$ and $G_2$ in a region
$-1\le\theta-\pi/2\le 1, -1\le\phi\le 1$ with $60\times 60$ sampling
points by taking the summation with respect to $\ell$ and $m$.  In our
calculation of images, we pick data points of $X^2+Y^2<1/4$ in the
observer's screen, which defines the aperture of our imaging system.
We applied the Tukey window to reduce unwanted aliasing originating
from the sharp cutoff of the aperture in discrete Fourier
transformation in a finite region.  Around the equatorial plane
$\theta\approx\pi/2$, difference between the spheroidal harmonics
$S_{\omega\ell m}$ and the spherical harmonics $Y_{\ell m}$ is not so
large; indeed, the relative difference between them is smaller than $0.1$
even for $\ell\ge 5$ with $a \omega=5$. Thus, we evaluate the sum
in $G$ by replacing $S_{\omega\ell m}$ with $Y_{\ell m}$ to reduce
computational time. We checked that the relative difference of $G$ is less than
1 \% for $a\omega=5$; hence, we expect that this replacement does not
produce much qualitative difference in the images.

%%%%%%%%%%%%%%%%%%%%%%%%%%%%%%%%%%%%%%%%%
\subsubsection{Images for $a=0$ (Schwarzschild case)} 
Figure~\ref{fig:image0om5a} shows the reflection and transmission
coefficients in the $(\ell,m)$-plane as well as $G_1$ and $G_2$ for
$\omega=5$.  As the black hole is spherically symmetric, the
reflection and transmission coefficients have no $m$ dependence. The
$\ell$ dependence of the reflection coefficient contains information
of the phase shift of waves scattered by the black hole.  For
$\ell\le \ell_*\approx 25$,
$\mathcal{R}_{\omega\ell m}\sim 0, \widetilde{\mathcal{T}}_{\omega\ell
  m}\sim 1$, and incoming waves from spatial infinity are perfectly
absorbed by the black hole. \anno{In wave optics, the photon sphere
  corresponds to a boundary between perfect absorption and perfect
  reflection in the $(\ell ,m)$-plane. In the eikonal limit, the
  boundary is represented as a relation between $\ell$ and $m$, which
  corresponds to a set of bounded photon orbits.} This critical value
$\ell_*/\omega\sim 5M$ corresponds to the size of photon sphere of the
Schwarzschild black hole. $|\mathcal{R}_{\omega\ell m}|$ and
$|\widetilde{\mathcal{T}}_{\omega\ell m}|$ satisfy the conservation
law
$|\mathcal{R}_{\omega\ell m}|^2+|\widetilde{\mathcal{T}}_{\omega\ell
  m}|^2=1$.  The correlation functions $G_1$ and $G_2$ on the
observer's screen show circular interference fringes. $G_2$ has a
sharp peak at the origin, which originated from the incoming radiation
from the cosmological horizon. The Fourier transformation of this peak
provides a nearly homogeneous background intensity of
images. The imaginary part of $G$ is zero for $a=0$.

%%% 
\begin{figure}[H]    
  \centering
  \includegraphics[width=0.9\linewidth,clip]{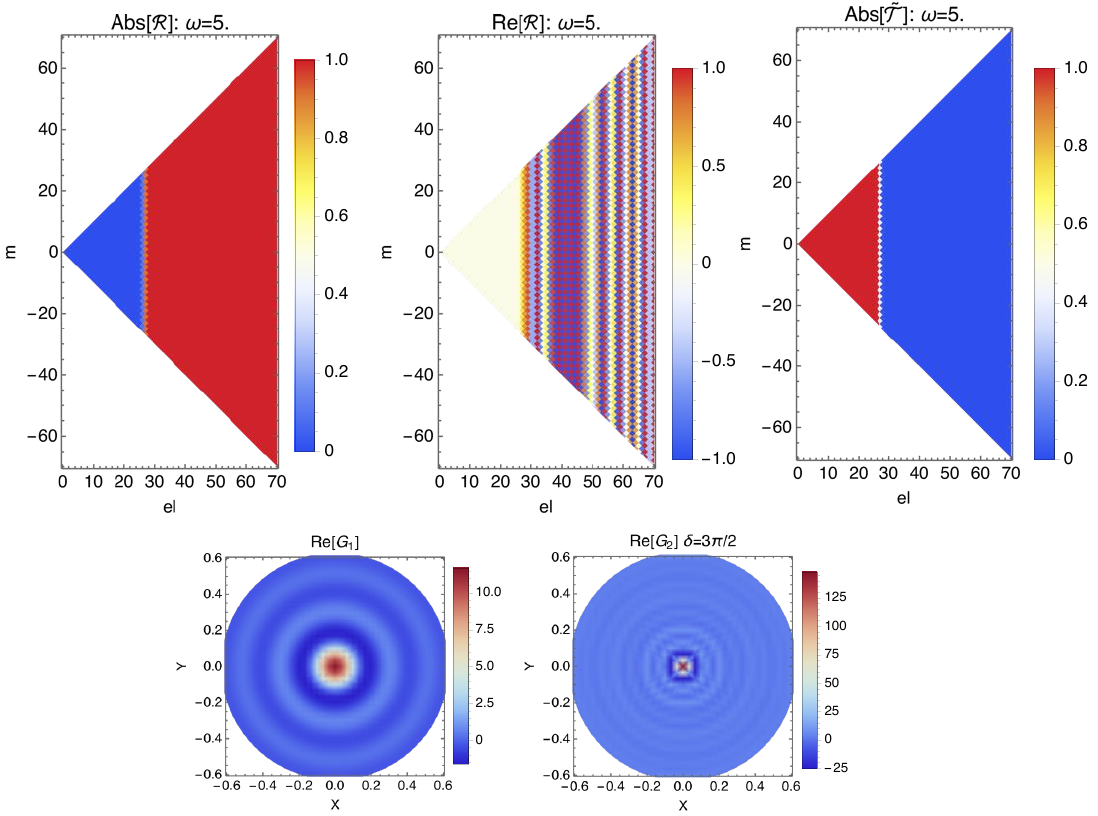}
  \caption{Reflection and transmission coefficients in the 
      $(\ell,m)$-plane (upper panels) as well as  $G_1$ and $G_2$ on the
    observer's screen (lower panels) for $a=0$ with $\omega=5$.  $G_1$
    shows interference fringes caused by Hawking radiation emitted by
    the black hole. $G_2$ shows the interference fringes caused by
    scattering of incoming radiation from the cosmological horizon by
    the black hole. A sharp peak in $G_2$ at $X=Y=0$ is due to the
    background incoming mode from the cosmological horizon, which
    results in homogeneous background intensity
    distribution in images. The imaginary part of $G_{1,2}$ is zero.}
  \label{fig:image0om5a}
\end{figure}
%%% 
Figure \ref{fig:image0om5b} shows images obtained by the Fourier
transformation of $G$.
%%%
$\mathcal{F}[G_1]$ is the image of the black hole illuminated by the
UP mode with the vacuum fluctuation. We superimpose the photon sphere
(dotted circle) with the image. As the photon sphere is a concept in
geometric optics, its shape has a finite width in wave optics. To
identify the location of the photon sphere in our calculation, we
define it as the location where the intensity of $\mathcal{F}[G_1]$
becomes half of that of the central bright region. The black hole has
the appearance of a ``shining star'' with its surface coinciding with
the photon sphere.  $\mathcal{F}[G_2]$ and
$\mathcal{F}[\widetilde G_2]$ are images of the black hole illuminated
by the IN mode, which is incoming radiation from the cosmological
horizon.  The emission from the black hole is not included in $G_2$ or
$\widetilde G_2$. The incoming radiation is scattered and absorbed by
the black hole. A dark circular shadow region surrounded by a bright
ring appears in these images. The shadow region is not black, because
$G_2$ has a contribution from incoming waves directly reaching the
detectors from the cosmological horizon (see Fig.~\ref{fig:toy}).
$\mathcal{F}[G_1+G_2]$ and $\mathcal{F}[G_1+\widetilde G_2]$ are
images with contributions from both the UP mode and the IN mode. The
black hole is visible as a bright disk in $\mathcal{F}[G_1+G_2]$. A
comparison of these two images reveals that the interference effect
sharpens the structure of the photon sphere. We can confirm this
behavior more clearly by checking one-dimensional slice of the images
(right panels of Fig.~\ref{fig:image0om5b}; slices of images along the
$Y_\text{im}=0$ line). Depending on the values of the phase $\delta$,
the intensity around the photon sphere becomes brighter or darker than
that of the image without the interference effect. The intensity
inside and outside of the photon sphere is not affected by the values of
$\delta$.
 
%%%
\begin{figure}[H]
  \centering
  \includegraphics[width=1.0\linewidth,clip]{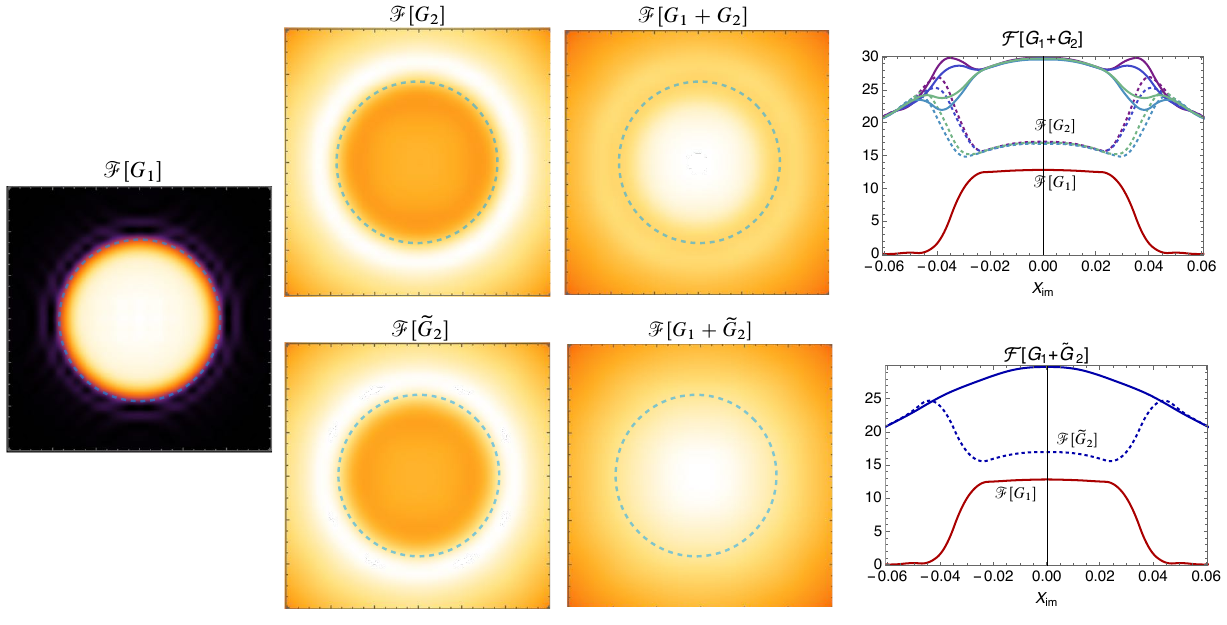}
  \caption{Images for $a=0$ and $\omega=5$. Two-dimensional images with
    $\del=3\pi/2$ and one-dimensional slice of images along
    $Y_\text{im}=0$ (right panels). Dotted circles in two-dimensional
    images represent the photon sphere. In the one-dimensional slice of
    $\mathcal{F}[G_1+G_2]$, images with four different phases
    $\del=0,\pi/2,\pi$, and $3\pi/2$, are shown.  }
  \label{fig:image0om5b}
\end{figure}
%%%
Figures~\ref{fig:image0om05a} and \ref{fig:image0om05b} show the
results for $\omega=0.5$.  From
$|\widetilde{\mathcal{T}}_{\omega\ell m}|$ and
$\mathcal{R}_{\omega\ell m}$, $\ell_{*}\approx 2$ corresponds to the
location of the photon sphere, but the shape of the photon sphere
becomes hazy for low frequency.  The image of $G_1$ spreads over the
field of view, and the structure of the photon sphere is not visible as an
image. This is because small $\ell$ modes mainly contribute to $G_1$
for the low-frequency case.  The image of $G$ shows that the entire field of
view becomes bright, and the brightness is much larger than that for
 $\omega=5$ because the emission of Hawking radiation is mainly
supported by the low-frequency mode $\omega\sim \kappa_{+}/(2\pi)$, 
the  wavelength of which  is much larger than the size of the photon sphere.

\newpage
%%%   
  \begin{figure}[H] 
  \centering
  \includegraphics[width=0.9\linewidth,clip]{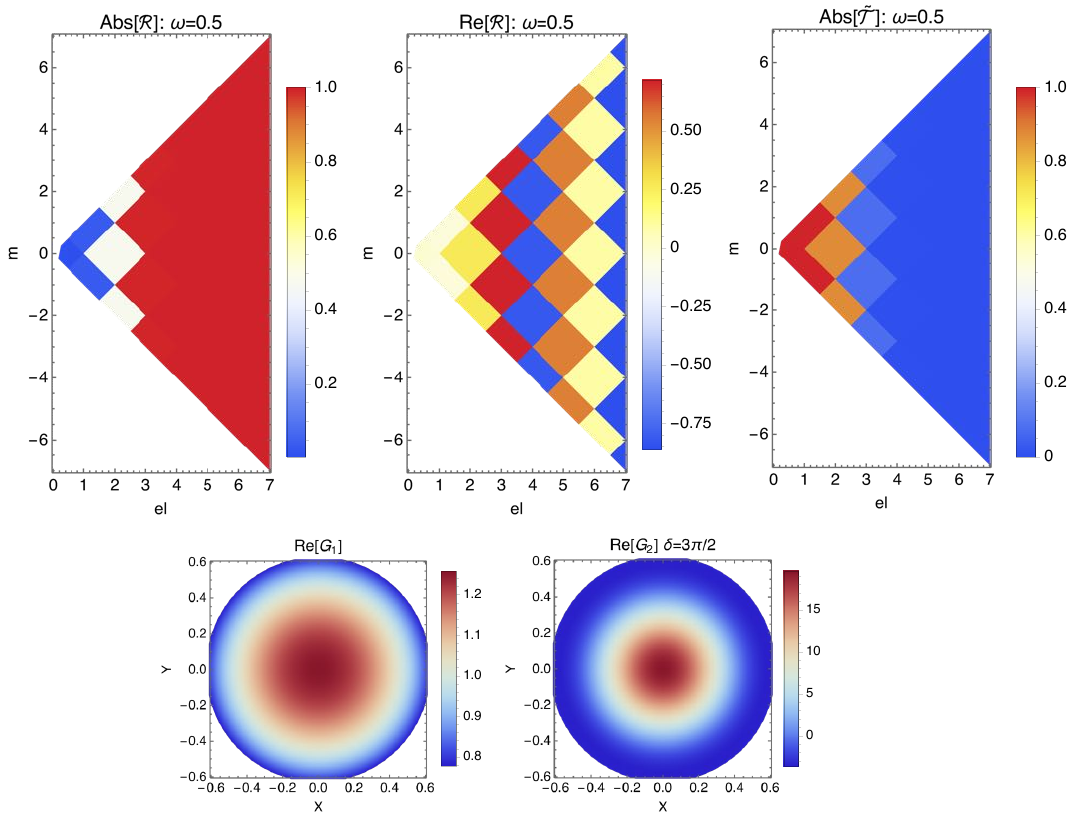}
  \caption{Reflection and transmission coefficients in the
    $(\ell,m)$-plane (upper panels) as well as $G_1$ and $G_2$ (lower
    panels) for $a=0$ and $\omega=0.5$.  The imaginary part of $G_{1,2}$ is
    zero.}
  \label{fig:image0om05a}
\end{figure}
%%%
\begin{figure}[H]
  \centering
  \includegraphics[width=0.9\linewidth,clip]{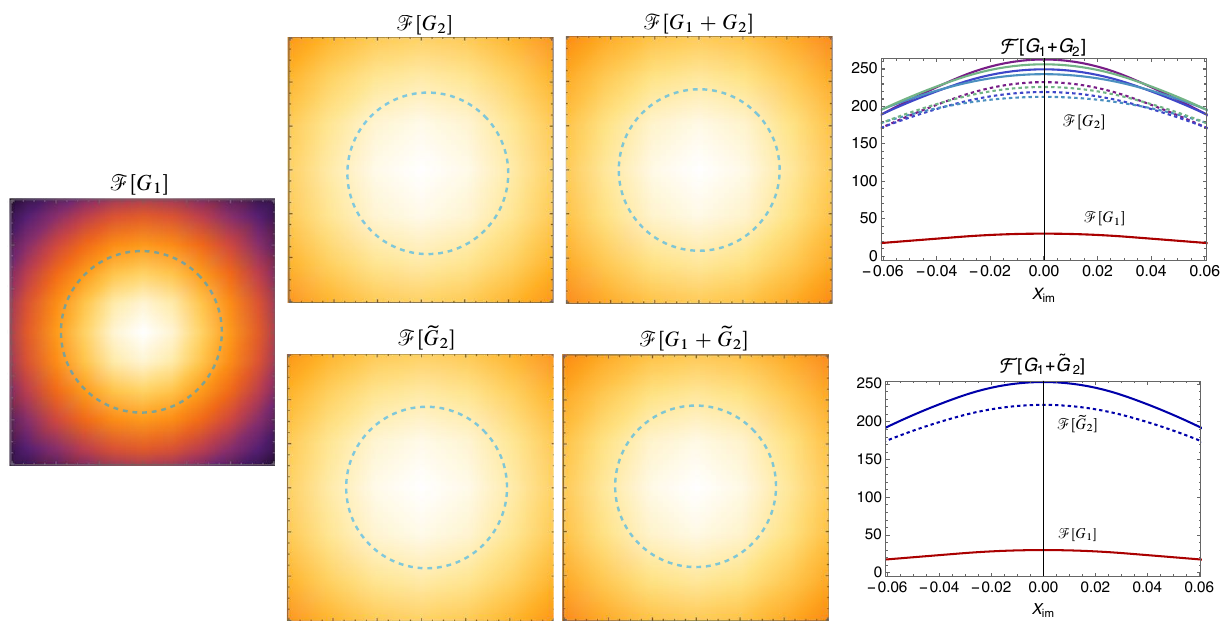}
  \caption{Images for $a=0$ and $\omega=0.5$. Two-dimensional images are
    drawn with $\del=3\pi/2$. Dotted circles in two-dimensional images
    represent the photon sphere. The entire field of view becomes
    bright, and it is not possible to identify the structure of the
    photon sphere.}
  \label{fig:image0om05b}
\end{figure} 
%%%%%%%%%%

\newpage
%%%%%%%%%%%%%%%%%%%%%%%%%%%%%%%%%%%%%%%%%%
\subsubsection{Images for $a=1/10$ (slowly rotating case)}
Figures~\ref{fig:image01om5a} and \ref{fig:image01om5b} show the
reflection and transmission coefficients as well as images for
$a=1/10$ and $\omega=5$.  The introduction of a small spin results in
a small deformation of $\mathcal{R}_{\omega\ell m}$ and
$\widetilde{\mathcal{T}}_{\omega\ell m}$ and causes the $m$ dependence
of these coefficients. The small spin parameter leads to a nonzero
imaginary part of $G_{1,2}$, which results in a left-right asymmetric
fringe pattern in the observer's screen.  The image of $G_1$ shows a
spherical photon sphere, which is not possible to distinguish for the
image of $a=0$. However, the images of $G_2$ shows an irregular-shaped
ring, which is caused by the interference effect between incoming and
outgoing waves. Indeed, the image of $\widetilde G_2$ shows a circular
 ring corresponding to the photon sphere. Moreover, the interference effect
enhances the left-right asymmetry of intensity of images around the photon
sphere, which is clearly visible from the one-dimensional slice of
images (right panels of Fig.~\ref{fig:image01om5b}).

%%% 
\begin{figure}[H] 
  \centering
  \includegraphics[width=0.9\linewidth,clip]{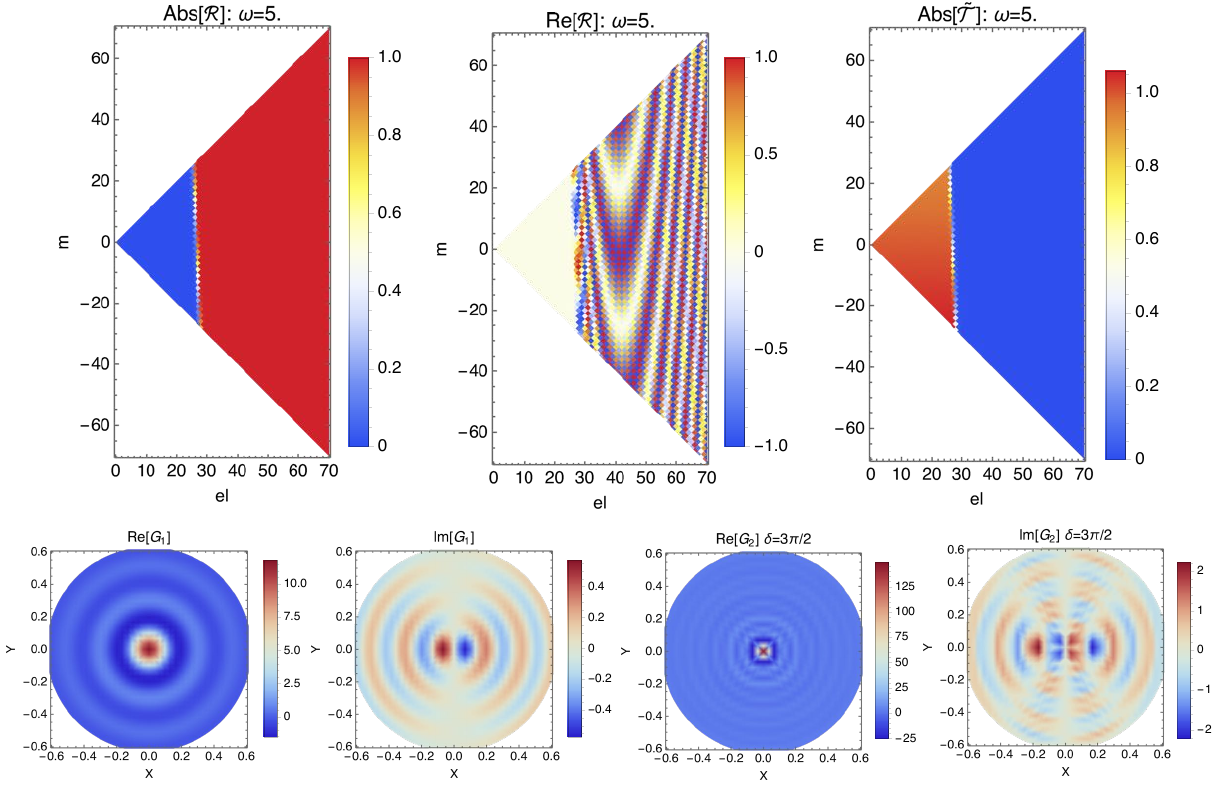}
  \caption{Reflection and transmission coefficients in the
    $(\ell,m)$-plane (upper panels) as well as $G_1$ and $G_2$ (lower
    panels) for $a=0.1$ and $\omega=5$. The reflection and transmission
    coefficients show $m$ dependence. The imaginary part of $G_{1,2}$
    presents left-right asymmetric fringe patterns, which are caused by
    the small value of the spin parameter.}
   \label{fig:image01om5a}
\end{figure}
%%%%
\begin{figure}[H]
  \centering
  \includegraphics[width=0.9\linewidth,clip]{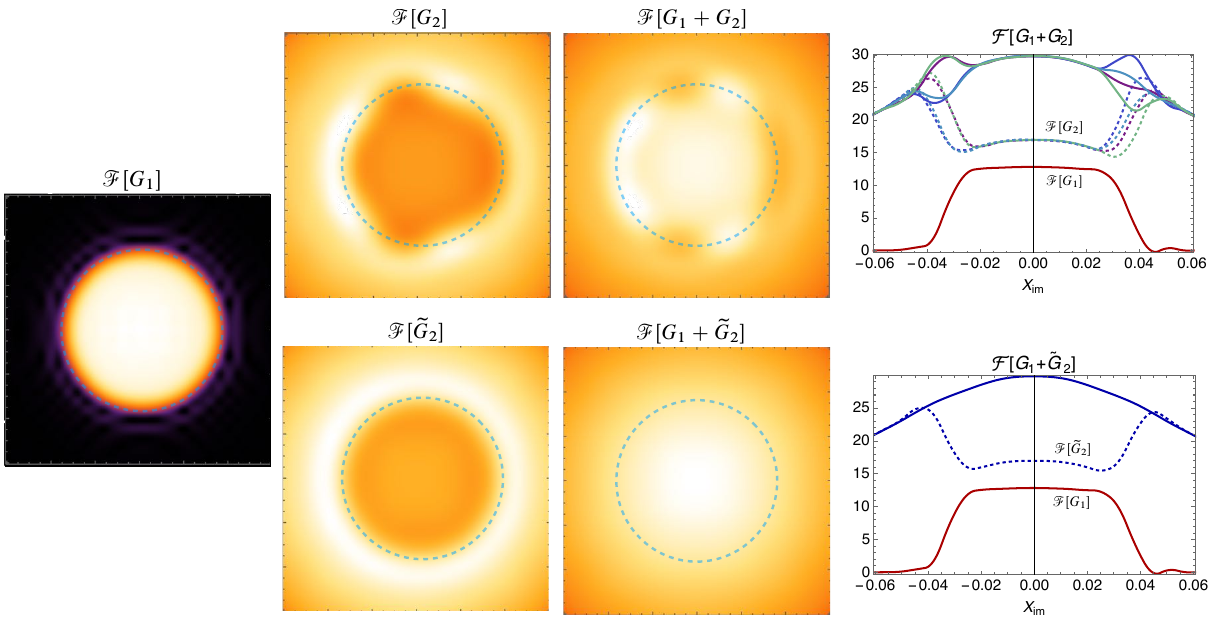}
  \caption{Images for $a=1/10$ and $\omega=5$. Two-dimensional images
    are obtained with $\del=3\pi/2$. Dotted circles in the
    two-dimensional images represent the photon sphere. The asymmetry of
    $\mathcal{F}[G_2]$ and $\mathcal{F}[G_1+G_2]$ reflect the effect of
    the small nonzero value of the spin parameter.}
  \label{fig:image01om5b}
\end{figure}
%%%

Figures~\ref{fig:image01om05a} and \ref{fig:image01om05b} show the
reflection and transmission coefficients as well as images for
$a=1/10$ and $\omega=0.5$.  $|\widetilde{\mathcal{T}}_{\omega\ell m}|$
and $\mathcal{R}_{\omega\ell m}$ show a small $m$ dependence. It is
not possible to distinguish the $\ell$ dependence from that for $a=0$
and $\omega=0.5$. The small $a$ induces the imaginary part of $G$,
which shows left-right asymmetry.  We cannot recognize the shape of
the emission region of Hawking radiation
(Fig.~\ref{fig:image01om05b}). The effect of the small spin parameter
is manifested as a small left-right asymmetry in the one-dimensional
intensity distribution (right panel of Fig.~\ref{fig:image01om05b}).
%%%
\begin{figure}[H]
  \centering
  \includegraphics[width=0.9\linewidth,clip]{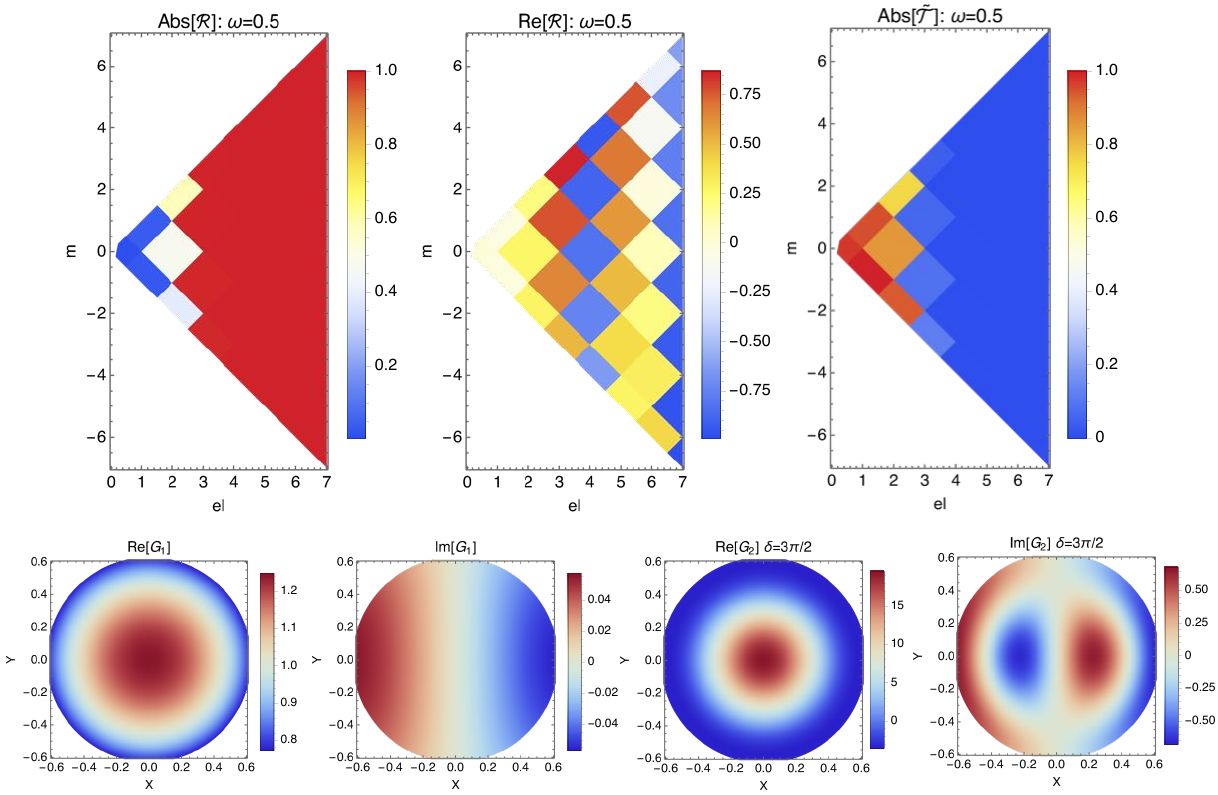}
  \caption{Reflection and transmission coefficients in the
      $(\ell,m)$-plane (upper panels) as well as $G_1$ and $G_2$ (lower panels)
    for $a=1/10$ and $\omega=0.5$. The left-right asymmetry of fringe
      patterns of $\mathrm{Im}[G_1]$ and $\mathrm{Im}[G_2]$ is also
      shown for the low-frequency case.}
  \label{fig:image01om05a}
\end{figure}
\begin{figure}[H]
  \centering
  \includegraphics[width=0.9\linewidth,clip]{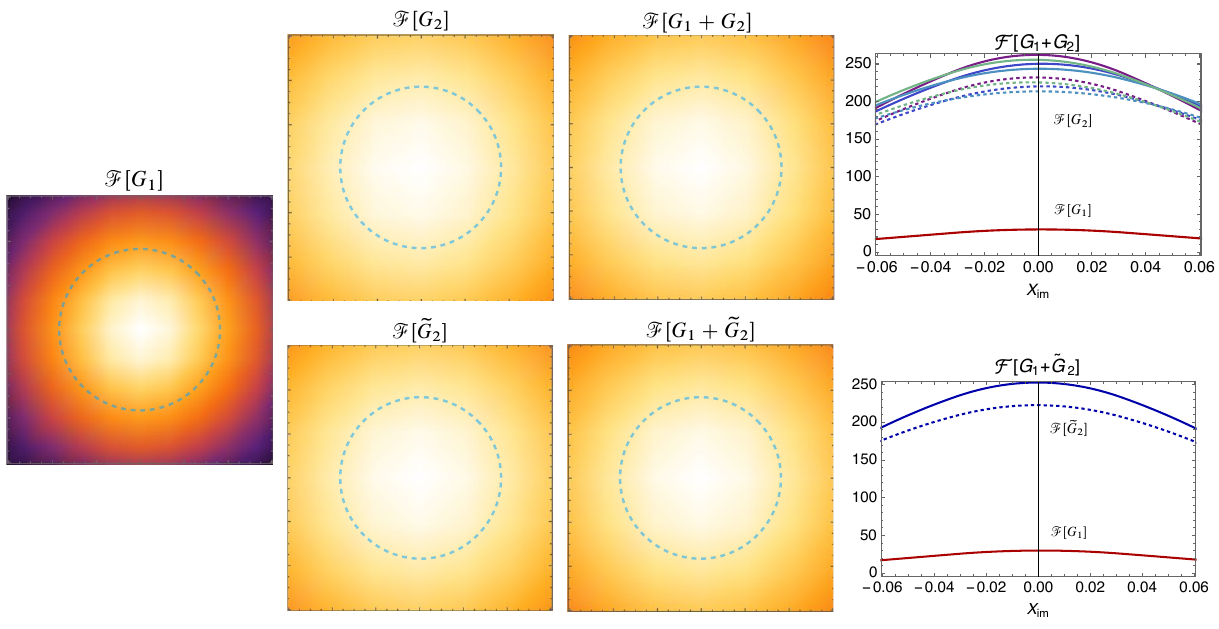}
  \caption{Images for $a=1/10$ and $\omega=0.5$. Two-dimensional
    images are obtained with $\del=3\pi/2$. Dotted circles in
    two-dimensional images represent the photon sphere. Left-right
    asymmetry is visible in $\mathcal{F}[G_1+G_2]$.}
  \label{fig:image01om05b} 
\end{figure}

%%%%%%%%%%%%%%%%%%%%%%%%%%%%%%%%%
\subsubsection{Images for $a=1$ (fast-rotating case)}
Figure~\ref{fig:image1om5a} shows the transmission and reflection
coefficients for $a=1$ and $\omega=5$.  The transmission and refection
coefficients show a larger $m$ dependence than those for $a=0.1$, and
the boundary between perfect reflection and absorption is much
deformed from an $\ell=\text{constant}$ line. This reflects the
non-spherical shape of the photon sphere. The interference fringe
pattern $\mathrm{Re}[G_1]$ is elongated in the $Y$ direction and
becomes elliptic owing to the spin of the black hole. Although
superradiant modes are included in the sum of the correlation
function, the impact of these modes on images is not visible because
the amplification factor is small for the scalar mode superradiance.

\begin{figure}[H]
  \centering
  \includegraphics[width=0.9\linewidth,clip]{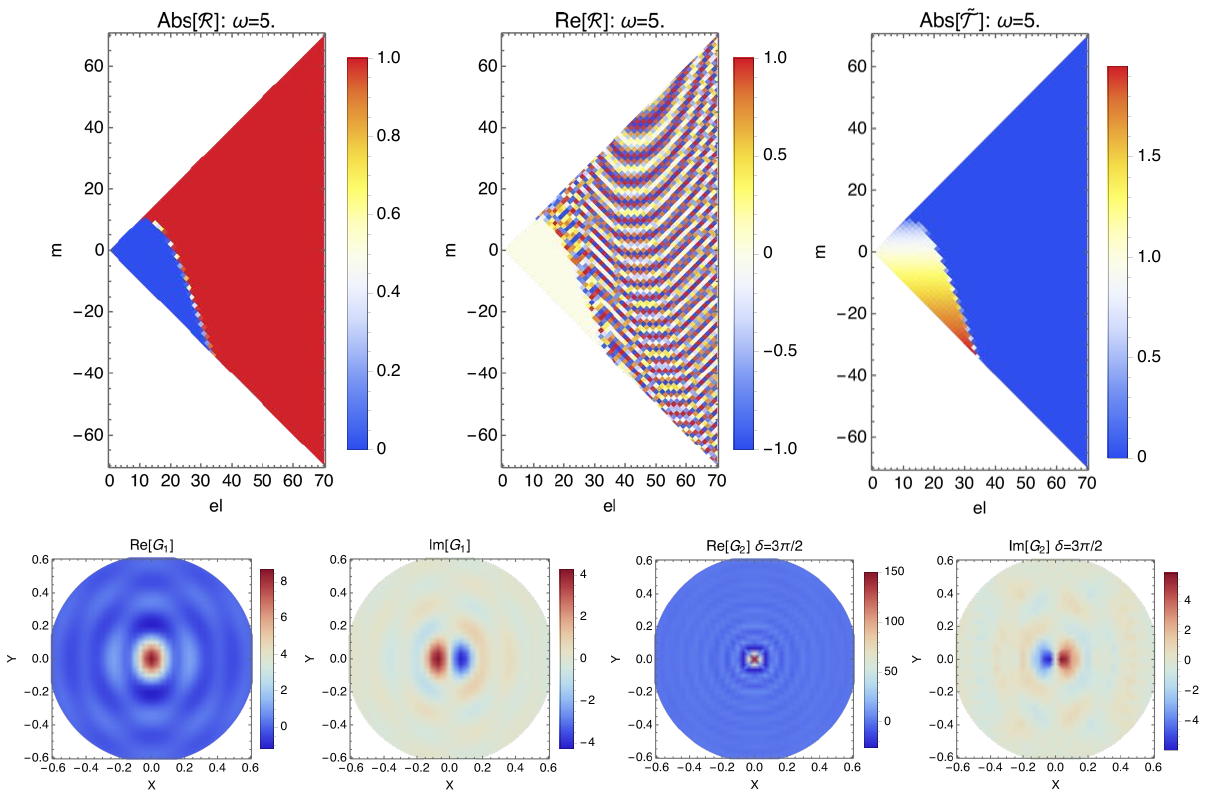}
  \caption{Reflection and transmission coefficients in the 
      $(\ell,m)$-plane (upper panels) as well as  $G_1$ and $G_2$ (lower panels)
    for $a=1$ and $\omega=5$. The interference fringe pattern
      $\mathrm{Re}[G_1]$ is elongated in the $Y$ direction and becomes
      elliptic because of the spin of the black hole.}
  \label{fig:image1om5a}
\end{figure}
%%%
Figure \ref{fig:image1om5b} shows images obtained from $G$.  The image
of $G_1$ shows a deformed D-shaped region corresponding to shape of
the photon sphere. The intensity inside of the photon sphere slightly
decreases as $X_\text{im}$ increases because of the dragging effect of
the Kerr black hole. The image of $G_2$ shows a dark shadow region
caused by absorption of the incoming radiation from the cosmological
horizon by the black hole. A peculiar feature of this image is a
bright spot at the left side of the photon sphere; this enhancement of
the intensity is due to the interference effect because we could not
find any intensity enhancement in the image of $\widetilde{G}_2$. As
we can see from the one-dimensional slice of images of
\anno{$\mathcal{F}[G_1+G_2]$}(right panels of
Fig.~\ref{fig:image1om5b}), the right side of the photon sphere can
also become bright depending on the value of $\del$. However, the
intensity \anno{of $\mathcal{F}[G_1]$} at the left side is larger than
that at the right side, and this difference is due to \anno{the
  dragging effect of the Kerr black hole: for
  $\pi\omega_{+}/\kappa_{+}\gg1$, the factor
  $\coth(2\pi\omega_{+}/\kappa_{+})|\widetilde{\mathcal{T}}_{\ell
    m}|^2/\omega_{+}$ in the summation of $G_1$ is approximated to
  $1/\omega_{+}=1/(\omega-m\,\Omega_{+})$. The $m$ dependence of this
  factor represents the left-right asymmetry of intensity in the image
  because in the eikonal limit, $m$ corresponds to the $z$ component
  of photon angular momentum and is related to the screen coordinate
  by $m/\omega\propto -X_\text{im}$~\cite{Fro2}. The intensity of the
  photon sphere projected on the screen is determined by this factor
  after mapping $m$ to the screen coordinate $X_\text{im}$.  A
  positive $m$ is mapped to a negative $X_\text{im}$, and a negative
  $m$ is mapped to a positive $X_\text{im}$. From the positive
  direction of $m$ (corotating direction) to the negative direction of
  $m$ (counter-rotating direction), this factor decreases because of 
  $\omega-m\,\Omega_{+}$ dependence, which is caused by the non-zero value
  of $\Omega_{_+}$.  }

\begin{figure}[H]
  \centering
  \includegraphics[width=0.9\linewidth,clip]{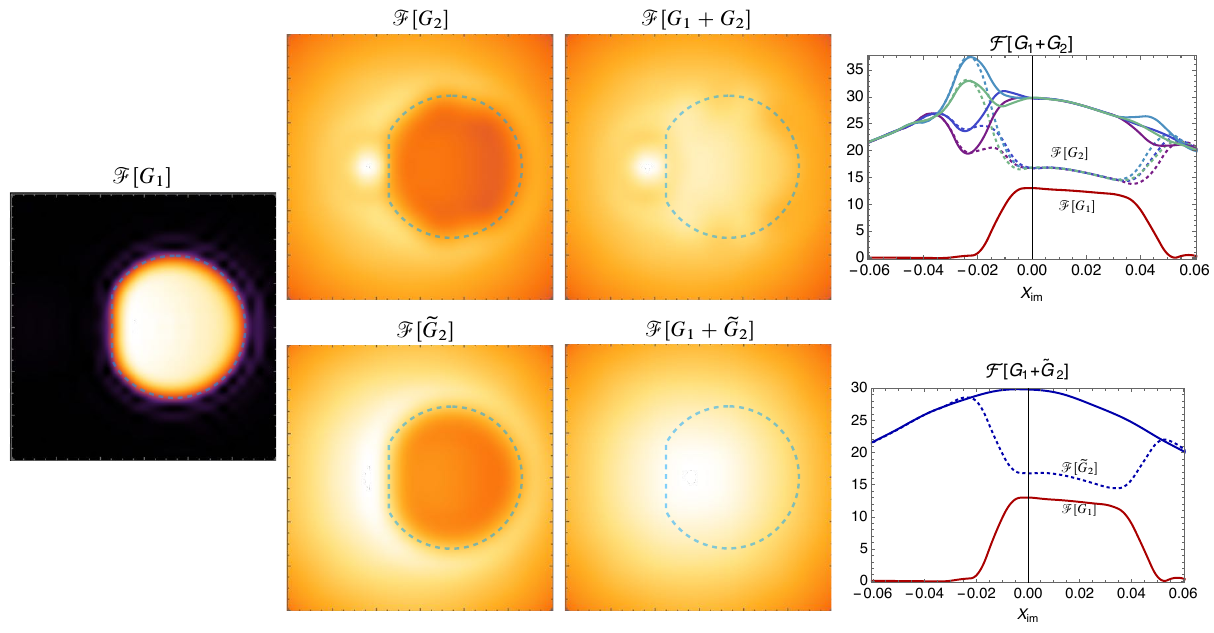}
  \caption{Images for $a=1$ and $\omega=5$. Dotted circles in
      two-dimensional images represent the photon sphere.
    $\mathcal{F}[G_2]$ and $\mathcal{F}[G_1+G_2]$ are images with
    $\del=3\pi/2$, and they show bright spots at the left side of the
    photon sphere, which are caused by interference between incoming
     and reflected modes.}
  \label{fig:image1om5b}
\end{figure} 
%%% 

Figures \ref{fig:image1om05a} and \ref{fig:image1om05b} show 
transmission and reflection coefficients as well as images for $a=1$
and $\omega=0.5$.
Although we cannot identify the photon sphere in the images, the left-right
asymmetry of intensity caused by the dragging effect of fast rotation
of the black hole is visible.
%%%
\begin{figure}[H]
  \centering
  \includegraphics[width=0.9\linewidth,clip]{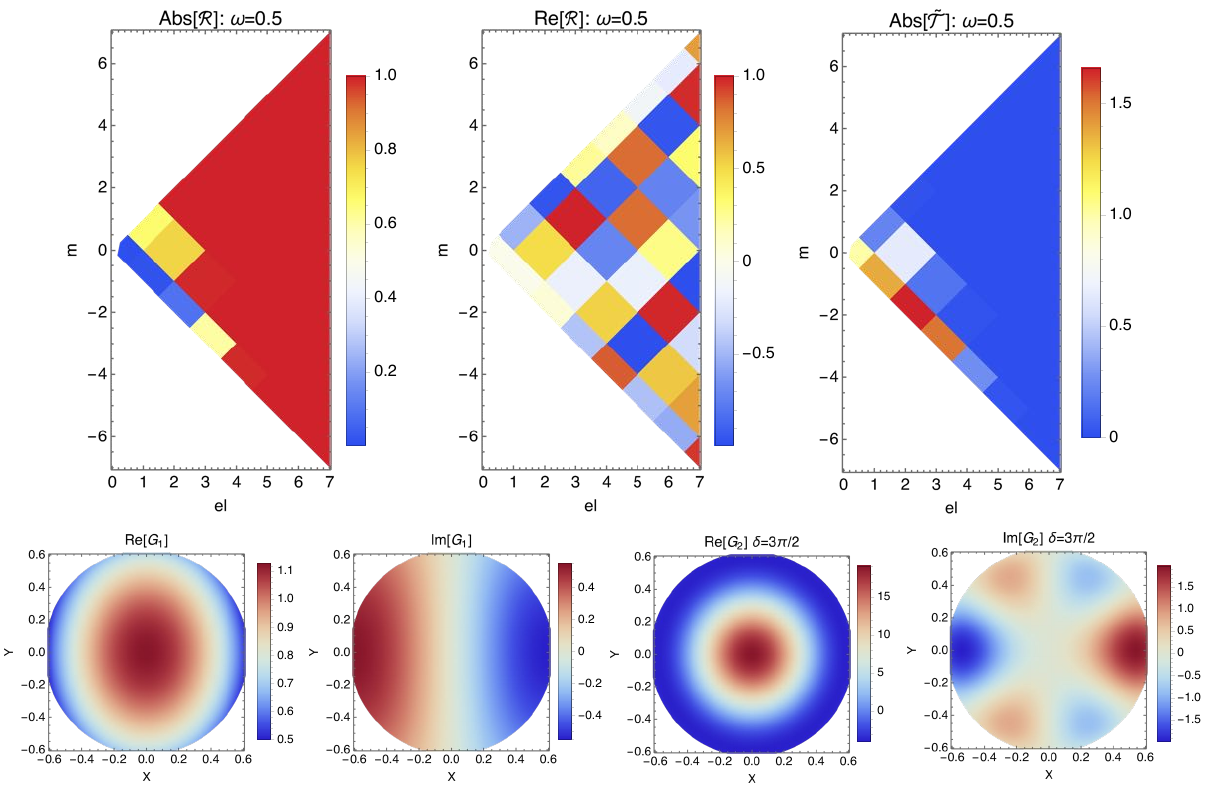}
  \caption{Reflection and transmission coefficients in the
    $(\ell,m)$-plane (upper panels) and $G_1$ and $G_2$ (lower panels)
    for $a=1$ and $\omega=0.5$. $\mathrm{Im}[G_{1,2}]$ shows
    left-right asymmetry due to the spin of the black hole.}
  \label{fig:image1om05a}
\end{figure}
\begin{figure}[H] 
  \centering
  \includegraphics[width=0.9\linewidth,clip]{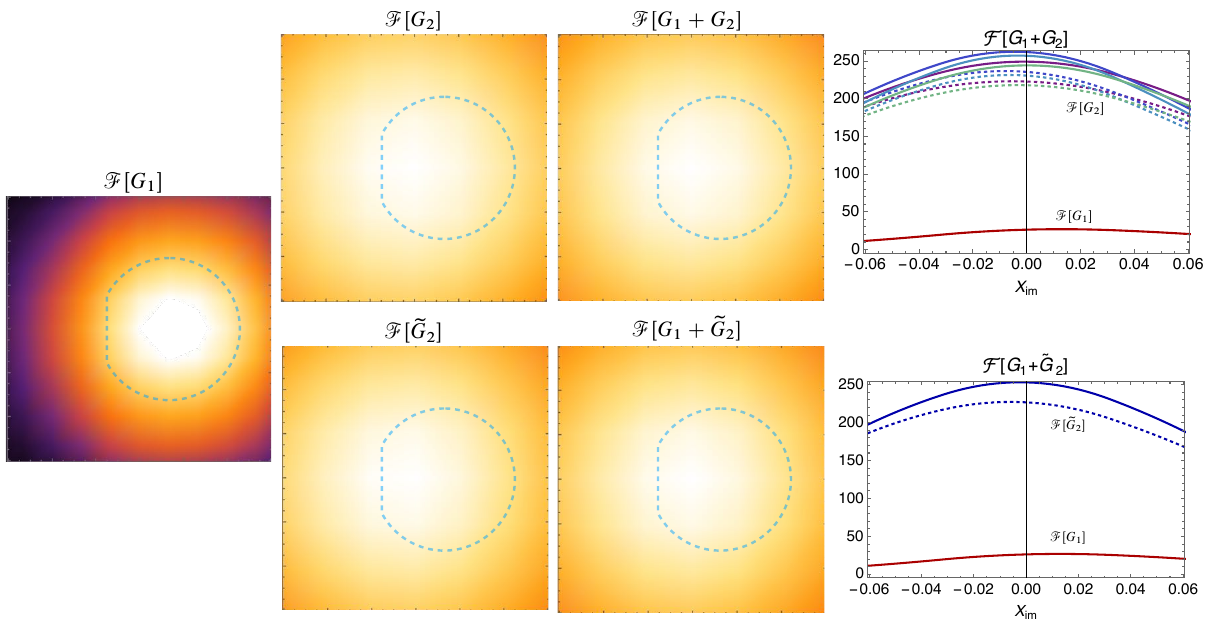}
  \caption{Images for $a=1$ and $\omega=0.5$. Two-dimensional images
    are obtained with $\del=3\pi/2$. Dotted circles in two-dimensional
    images represent the photon sphere.  $\mathcal{F}[G_2]$ and
    $\mathcal{F}[G_1+G_2]$ are images with $\del=3\pi/2$. Left-right
    asymmetry is visible in the one-dimensional slice of
    $\mathcal{F}[G_1+G_2]$.}
  \label{fig:image1om05b} 
\end{figure}
%%%%%%%%%%%%%%%

\subsubsection{Images of $G_1$ and  emission region of Hawking radiation}
Figure~\ref{fig:slice} shows the one-dimensional slice of images
$\mathcal{F}[G_1]$ with different frequencies. For $\omega=5$, we can
identify the location of the photon sphere, which is visible as sharp
edges in the image. The size of the effective emission area of Hawking
radiation is the same as that of the photon sphere, as discussed
in~\cite{Giddings2016,Dey2017a}. For low frequency, the effective size
of the radiation source obtained from the images becomes larger than the
photon sphere. This is because the $\ell=0$ mode mainly contributes to the
greybody factor $|\widetilde T_{\omega\ell m}|^2$ for
$\omega\rightarrow 0$ and the characteristic size of the emission region
depends on $\omega$. The introduction of the spin of the black hole does
not alter this behavior of $\omega$ dependence of the size of the emission
region of Hawking radiation.
%%%%
\begin{figure}[H]
  \centering
  \includegraphics[width=1.\linewidth,clip]{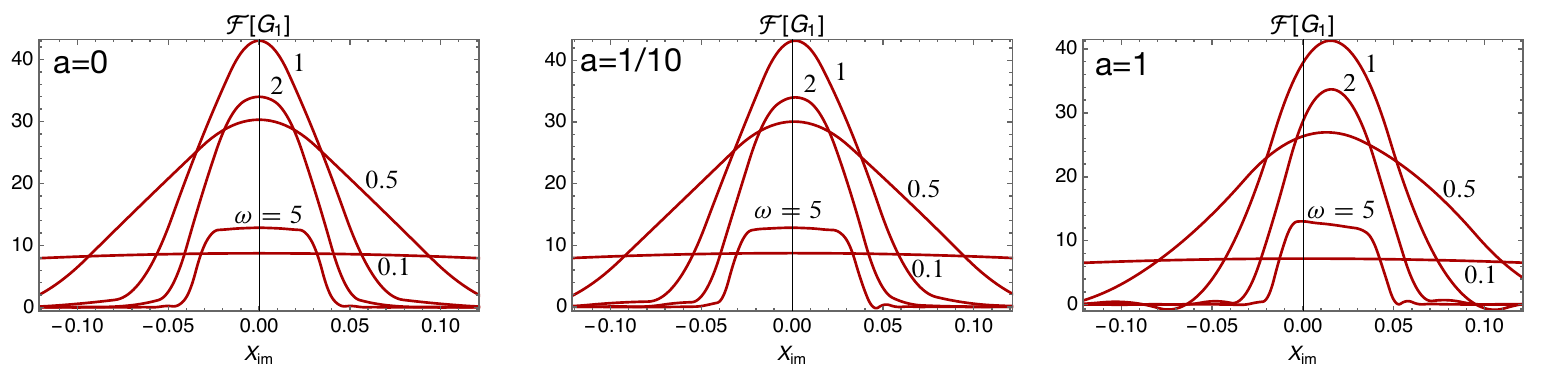}
  \caption{Slices of images $\mathcal{F}[G_1]$ along the
    $Y_\text{im}=0$ line. For lower frequencies, the effective size of
    emission region becomes larger than the photon sphere.}
  \label{fig:slice}
\end{figure} 
%%%
For obtaining black hole images directly related to the emission of
Hawking radiation, we consider images of the UP mode with the
correlation function $G_1-G_1^{\text{Boulware}}$ given by
Eq.~\eqref{eq:G-planck}, which subtracts contribution of the vacuum
fluctuation (Fig.~\ref{fig:image-planck}). The obtained images are
sensitive to the values of the spin parameter. For $a\neq 0$, owing to
the $m$-dependence of the Planckian factor in \eqref{eq:G-planck}, 
\anno{
  $(e^{2\pi\omega_{+}/\kappa_{+}}-1)^{-1}|\widetilde{\mathcal{T}}_{\ell
    m}|^2/\omega_{+}\sim e^{-2\pi\omega_{+}/\kappa_{+}}/\omega_{+}$
  for $\pi\omega_{+}/\kappa_{+}\gg 1$,} the intensity of the emission
region around the left side of the photon sphere becomes larger and
decays exponentially while departing from this region. The peak intensity of
the emission region strongly depends on the spin parameter:
$9.5\times 10^{-49}$ for $a=0.1$ and $4.8\times 10^{-10}$ for
$a=1$. As this $m$ dependence in the Planckian factor is proportional
to the angular velocity of the black hole, the images of
$G_1-G_1^\text{Boulware}$ reflect the dragging effect in the vicinity
of the event horizon of the Kerr black hole.

\begin{figure}[H]
  \centering 
  \includegraphics[width=1\linewidth,clip]{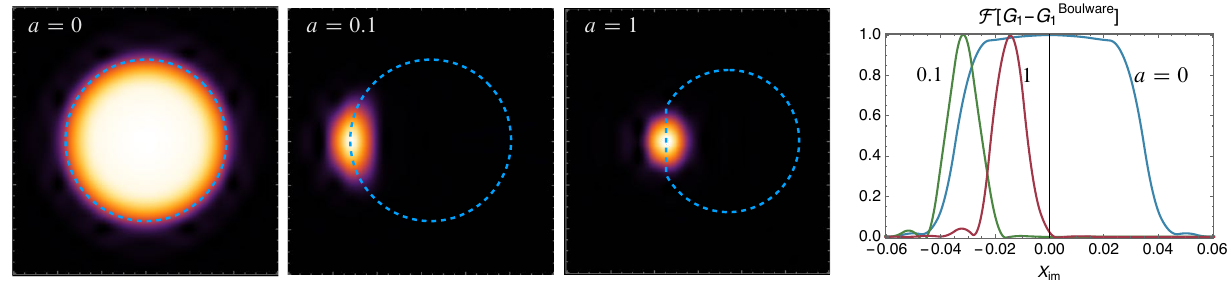}
  \caption{Images of the UP mode for $G_1-G_1^\text{Boulware}$ with
    $\omega=5$. Dotted circles indicate the photon sphere. The
    right panel is the one-dimensional slice of images along
    $Y_\text{im}=0$. The intensity is normalized by its peak
    value. The unnormalized values of the intensity are
    $5.1\times 10^{-57}$ for $a=0$, $9.5\times 10^{-49}$ for $a=0.1$ and
    $4.8\times 10^{-10}$ for $a=1$.}
  \label{fig:image-planck}
\end{figure}
%%%%%%%%%%%%%%%%%%%%%%%%%%%%%%%%%%%%%

\section{Summary}

We investigate the wave optical imaging of black holes using Hawking
radiation.  For the theoretical investigation of the imaging of
astrophysical black holes, Falcke \textit{et al.}~\cite{Falcke2000}
obtained images of black hole shadows using the ray tracing of photons
emitted from infalling gas around a Kerr black hole. Their images show
a left-right asymmetry of intensity due to the spin of the black hole
and frequency dependence of images such that the ``shadow'' becomes
invisible for low frequency because of the scattering of photons by
the plasma around the black hole. Comparing their images with those
obtained in the present study, the apparent structure of our images
$\mathcal{F}[G_2]$ (Figs.~\ref{fig:image1om5b} and
\ref{fig:image1om05b}) resembles theirs. However, images
$\mathcal{F}[G_1+G_2]$ show no specific structure associated with the
photon sphere. As discussed in Section \ref{sec:simple-model}, for
high-frequency waves beyond the Hawking temperature, the structure of
the images $\mathcal{F}[G_1+G_2]$ becomes flat if we neglect the
interference effect. In our calculation, the adopted frequency is far
higher than the Hawking temperature with 
$\omega_{+}\approx 0.006, \omega_c\approx 0.008$ for $a=1$, which is
why we do not have shadow images for $\mathcal{F}[G_1+G_2]$. Although
it is not possible to identify the exact location of the wave source for
Hawking radiation, we applied the van Cittert-Zernike theorem and
obtained projected two-dimensional images of the black hole using
the Fourier transformation of spatial correlation functions.  The obtained
images trace the shape of the photon sphere of the black hole for high
frequency, and  the black hole has appearance of  a shining star with
the photon sphere as its surface.  For low frequency, a definite
surface of emission is lost, and the emission region extends over entire
field of view and is larger than  the photon sphere. We found that
interference between incoming modes from the cosmological horizon and
 modes reflected by the black hole enhances the intensity of images in the
vicinity of the photon sphere for fast-spinning black holes.

We are not certain whether the ``source'' region of Hawking radiation
is spatially incoherent, which is a crucial assumption of the van
Cittert-Zernike theorem for imaging. However, although the detail of
the spatial incoherence is not justified, it is possible to adopt the
Fourier transformation of the spatial correlation function as a tool
to visualize black holes with Hawking radiation. The spatial
correlation function of the UP mode near the black hole (near the past
event horizon) is roughly estimated as follows. The radial wave
function in the vicinity of the horizon is
$R_{\ell m}^\text{up}\sim e^{-i(\omega-m\Omega)r_*}$, and
%%% 
\begin{equation}
    G_1(\theta_1,\phi_1,\theta_2,\phi_2)
    \sim\sum_{\ell m} |R^\text{up}_{\ell m}(r)|^2S_{\ell
      m}(\theta_1)S_{\ell
      m}(\theta_2)e^{im(\phi_1-\phi_2)}\sim\delta(\theta_1-\theta_2)\del(\phi_1-\phi_2)
\label{eq:G-hor}
\end{equation}
%%%
because $|R_{\omega\ell m}^\text{up}|^2$ does not have an $\ell$ or
$m$ dependence. Therefore, if we assume that Hawking radiation is
emitted from an $r=$constant surface in the vicinity of the horizon,
the spatial coherence of that source surface is zero and we have
justified the applicability of the van Cittert-Zernike theorem to imaging
with Hawking radiation. The emission region of Hawking radiation may
differ from the vicinity of the horizon, and for such a case, we cannot
make assertions on the spatial incoherence of the source region of
Hawking radiation. The spatial coherence of the source will result in hazy
images, and more rigorous investigation on the spatial coherence of Hawking
radiation will reveal its ``quantumness''.  The word ``quantumness''
is obscure, and we should properly define it based on
entanglement.  If we reconsider the van Cittert-Zernike theorem for a
source with spatial coherence, it may be possible to access
information on the degree of coherence of Hawking radiation.  This
direction of investigation is related to entanglement harvesting in
black hole spacetimes with the method of intensity
correlation~\cite{Baym1998}, and we will report on this subject in a
separate publication.

%%%%%%%%%%%%%%%%%%%%%%%%%%%%%%%%%%%%%%%%%%%%%%%%
%\acknowledgements{
%Y.N. was supported in part by JSPS KAKENHI Grant No. 19K03866. }

\acknowledgements
Y.N. was supported in part by JSPS KAKENHI Grant No. 19K03866. 

\anno{
\appendix
\section{Van Cittert-Zernike theorem in de Sitter spacetime}
In this Appendix, we consider the van Cittert-Zernike theorem in de
Sitter spacetime with the metric
%%%
\begin{equation}
  ds^2=-fdt^2+\frac{dr^2}{f}+r^2d\Omega^2,\quad
  f=1-\frac{\Lambda}{3}r^2,\quad \Lambda>0.
\end{equation}
%%%
A massless conformal scalar field $\Phi$ obeys
%%%
\begin{equation}
  \left(\square-\frac{R}{6}\right)\Phi=0,\quad R=4\Lambda,
\label{eq:SGreen}
\end{equation}
%%%
and the scalar field is separated as
%%%
\begin{equation}
  \Phi=e^{-i\omega t}\frac{R_\ell(r)}{r}Y_{\ell m}(\theta,\phi).
\end{equation}
%%%
The radial wave equation is
%%%
\begin{equation}
  \left[\frac{d^2}{dr_*^2}+\omega^2-\frac{\ell(\ell+1)H^2}{\sinh^2(Hr_*)}\right]R_\ell(r_*)=0,\quad
  Hr=\tanh(Hr_*),\quad H^2=\frac{\Lambda}{3},
\end{equation}
%%%
where $r_*=\int dr/f$ is the tortoise coordinate.  The Green's function
for Eq.~\eqref{eq:SGreen} is given by
%%%
\begin{equation}
  \mathcal{G}_\omega(\bs{x},\bs{x}_s)=\frac{i\omega}{4\pi rr_s}\sum_{\ell
    }R^{(1)}_\ell(r_{s*})R^{(2)}_\ell(r_*)(2\ell+1)P_\ell(\bs{n}\cdot\bs{n}_s),
\end{equation}
%%%
where $\bs{x}=r_*\bs{n}$ and $\bs{x}_s=r_{s*}\bs{n}_s$ with
$|\bs{n}|=|\bs{n}_s|=1$, and $R_\ell^{(1)}$ is regular at $r=0$, and
$R^{(2)}_\ell$ is outgoing at $r_*=\infty$. We assume $\ell/\omega\ll
1/H$, which means that the ``impact parameter'' of the wave mode
with $\ell$ is smaller
than the Hubble horizon length and that the effect of the cosmological constant
is negligible. Under this condition, for $r_{s*}\ll 1/(\omega
H)$, $R^{(1)}_\ell(r_{s*})\sim r_{s*}j_\ell(\omega r_{s*})$, and for $1\ll \omega
r_*$, $R^{(2)}_\ell(r_*)\sim (-i)^{\ell+1}e^{i\omega
  r_*}/\omega$. Thus, the Green's function behaves as
%%%
\begin{align}
  \mathcal{G}_\omega(\bs{x},\bs{x}_s)
  &\propto\frac{e^{i\omega r_*}}{r}\sum_\ell(2\ell+1)(-i)^\ell
    j_\ell(\omega r_{s*})P_\ell(\bs{n}\cdot\bs{n}_s) \notag \\
  &=\frac{1}{r}\exp\bigl[i\omega(r_*-\bs{x}\cdot\bs{x}_s/r_*)\bigr].
\end{align}
%%%
This Green's function corresponds to \eqref{eq:Green} for the case of
flat space. The only difference is that the radial coordinate is
replaced with the corresponding tortoise coordinate. Therefore, the
van Cittert-Zernike theorem for de Sitter spacetime has the same form
as the flat case with the replacement $r\rightarrow r_*$ in the phase
factor.  }

%\bibliography{000:My_projects,book,001:quantum_physics,sonic_analog}
%\bibliography{cosmology,my-paper,relativity,quantum,black-hole}

\end{document}